\documentclass[fleqn,usenatbib]{mnras}

\usepackage{newtxtext,newtxmath}

\usepackage[T1]{fontenc}

\DeclareRobustCommand{\VAN}[3]{#2}
\let\VANthebibliography\thebibliography
\def\thebibliography{\DeclareRobustCommand{\VAN}[3]{##3}\VANthebibliography}
\usepackage{graphicx}	
 
\usepackage{amsmath}	
\usepackage{amssymb}	
\usepackage{xfrac,nicefrac}

\newcommand{\Mdot}{\mbox{\,$\rm M_{\odot}$}}        
\newcommand{\Zdot}{\mbox{\,$\rm Z_{\odot}$}}        
\newcommand{\ZLMC}{\mbox{\,$\rm Z_{\rm LMC}$}}        
\newcommand{\ZSMC}{\mbox{\,$\rm Z_{\rm SMC}$}}        
\newcommand{\kms}{\,km s$^{-1}$}   			
\newcommand{\aov}{$\alpha_{\rm ov}$}		
\newcommand{\fov}{$f_{\rm ov}$}		

\title[Wolf-Rayet mass-loss rates]{Evolution of Wolf-Rayet stars as black hole progenitors}

\author[E. R. Higgins et al.]{
E. R. Higgins,$^{1}$\thanks{E-mail: erin.higgins@armagh.ac.uk}
A. A. C. Sander,$^{1}$
J. S. Vink,$^{1}$
and R. Hirschi$^{2,3}$
\\
$^{1}$Armagh Observatory and Planetarium, College Hill, Armagh BT61 9DG, N. Ireland\\
$^{2}$Astrophysics Group, Keele University, Keele, Staffordshire ST5 5BG, UK
$^{3}$Kavli Institute for the Physics and Mathematice of the Universe (WPI),\\ University of Tokyo, 5-1-5 Kashiwanoha, Kashiwa 277-8583, Japan}

\date{Accepted XXX. Received YYY}

\pubyear{2021}

\begin{document}
\label{firstpage}
\pagerange{\pageref{firstpage}--\pageref{lastpage}}
\maketitle

\begin{abstract}
Evolved Wolf-Rayet stars form a key aspect of massive star evolution, and their strong outflows determine their final fates.
In this study, we calculate grids of stellar models for a wide range of initial masses at five metallicities (ranging from solar down to just 2\% solar).
We compare a recent hydrodynamically-consistent wind prescription with two earlier frequently-used wind recipes in stellar evolution and population synthesis modelling, and
we present the ranges of maximum final masses at core He-exhaustion for each wind prescription and metallicity $Z$.
Our model grids reveal qualitative differences in mass-loss behaviour of the wind prescriptions in terms of "convergence".
Using the prescription from Nugis \& Lamers the maximum stellar black hole is found to converge to a value of 20-30\Mdot, independent of host metallicity, however
when utilising the new physically-motivated prescription from Sander \& Vink there is no convergence to a maximum black hole mass value.
The final mass is simply larger for larger initial He-star mass, which implies that the upper black hole limit for He-stars below the pair-instability gap is set by prior evolution with mass loss, or
the pair instability itself.
Quantitatively, we find the critical $Z$ for pair-instability (Z$_{\rm{PI}}$) to be as high as  50\% \Zdot, corresponding to the host metallicity of the LMC. 
Moreover, while the Nugis \& Lamers prescription would not predict any black holes above the approx 130\Mdot\ pair-instability limit, with Sander \& Vink winds included, we
demonstrate a potential channel for very massive helium stars to form such massive black holes at $\sim$ 2\% \Zdot\ or below.
\end{abstract}


\section{Introduction}
The most massive stars are central to many fields of Astrophysics, driving the evolution of galaxies, recycling fusion material, as well as enriching their surroundings through strong stellar winds. While they become increasingly rare with increasing mass, they are also the progenitors of the heaviest stellar mass black holes (BHs). Recently, the detection of merging BHs via gravitational waves (GWs) has queried the spectrum of BH masses for different galaxies, with massive BHs initially detected at 30-40\Mdot\ by \cite{Abbott+2016}. More recently, the detection of an 85\Mdot\ and 66\Mdot\ BH merger in GW\,190521 witnessed an interesting challenge for stellar evolution and GW theorists. The mass of each BH in this event appears to lie in the so-called 'pair instability' (PI) mass gap, where the star is thought to rip itself apart in a violent explosion or through pulsations in a `pulsational pair-instability' (PPI) supernova. As such, \cite{Abbott2020} suggested this extreme GW event to be the product of second generation BHs. Stellar evolution studies such as \cite{vink2020,farrell2021} have however hypothesised alternative solutions for creating first generation heavy BHs such as in GW\,190521 by maintaining a low core mass and retaining the large H envelope, which might even be possible for metallicities up to 0.1 \Zdot\ \citep{vink2020}.

The various evolutionary paths a massive star may take on route to becoming a black hole include stars which are stripped of their envelope at the end of core H-burning towards a Helium (He) star. For a sufficiently high mass-loss rate \citep{Sander+2020,Shenar2020}, such stars will appear as classical Wolf-Rayet (WR) stars. The star may have been stripped through binary interactions \citep[e.g.][]{Podsi92} or as a result of strong stellar winds in the case of single massive stars \citep[e.g.][]{yoon2012,georgy2013}. Massive He stars and WR stars are considered to be the final evolutionary stage before forming a BH, and as a result, WR mass loss is the key in establishing the BH progenitor mass range. 

In spite of this, the levels of He star mass-loss rates have been debated in particular with regards to uncertainties in their mass- and metallicity-dependence. For lower-mass objects, which will likely not appear as WR stars \citep{Vink2017,SV2020}, there is essentially an absence of observations already at solar metallicity ($Z = Z_\odot$) between hot subdwarfs \citep[e.g.][]{Schootemeijer+2018,Wang+2018} and WR stars \citep[e.g.][]{Hamann+2019,Sander19} with the notable exception of HD 45166 \citep{Groh+2008}. At low $Z$, the situation gets even worse due to the lack of observed WR stars below the regime of the Small Magellanic Cloud ($Z \approx 0.2\,Z_\odot$). Given the need to estimate the mass loss of He stars in stellar evolution, population synthesis, or supernova statistics \citep[e.g.][]{ES2009,Woosley+2020}, there is a high demand for theoretically determined mass-loss rates of massive stars, in particular for He stars and WRs.

Besides directly determining the upper-mass limit of BHs, the mass-loss dependent evolution of He stars as a function of metallicity also has consequences for the fraction of type IIb and Ibc stripped supernovae (SNe) as stronger winds could remove more material from the outer layers of a star, thereby altering the parameters of the eventual core collapse event \citep[e.g.][]{Gilkis+2019}. All this of course has major implications for the field of GW astronomy and detections of future BH mergers. 

Until 2005, those stellar evolution models that included a metallicity dependence of WR stars usually relied on the {\it total} metallicity \citep[e.g.][]{NugisLamers} which is dominated by self-enriched surface material (nitrogen in the case of WN stars, carbon for WC stars). In a pilot study of late-type WN and WC stars, \citet{Vink05} showed that it is not the CNO abundances but the \textit{initial} iron (Fe) abundance that is the key in setting the mass-loss metallicity dependence, enabling \citet{EV2006} to model the observed drop in the WC/WN ratio towards lower host galaxy metallicity, and to predict heavier BHs in lower $Z$ host galaxies. 

This initial $Z$, Fe dependent mass-loss of WR stars was also included in the GRB progenitor evolution by \cite{Yoon06} as well as in the maximum BH population study of \citet{Belczynski+2010}, which was subsequently utilised to infer the low host $Z$ of the first gravitational wave source GW\,150914 \citep[][]{Abbott+2016, vink19}. It is thus clear that in order to correctly predict the BH mass distribution with $Z$, we need to properly understand the intricacies of WR mass loss as a function of $Z$, appreciating there is no requirement for this dependence to be a simple power law. In fact, \citet{SV2020} recently showed a more complex behaviour, and it is our aim in the current paper to explore the implications of this new WR mass-loss formulations for the BH mass function with $Z$. We defer a more detailed study of WR evolution in terms of WN, WC, WO sub-types to the future.

In Sect. \ref{SectModels}, we present our method of evolving He stars and outline our model grid. Mass-loss rates of He stars are discussed in Sect. \ref{MasslossSect}, including the implementation of three comparative prescriptions in our evolution models. Results from our Standard Grid of models are presented in Sect. \ref{newgridSect}, while comparable results for our Alternative Grid of models are outlined in Sect. \ref{QCHEgrid}. A discussion of the upper mass limit of black holes in terms of pair instability is shown in Sect. \ref{PPI}, with conclusions outlined in Sect. \ref{conclusions}.

\section{Stellar evolution models}\label{SectModels}
We utilised the stellar evolution code MESA \citep[version 8845;][]{Pax11, Pax13, Pax15}, in calculating detailed models of massive stars, focusing on the core He-burning phase of evolution. We provide a range of initial helium Zero-Age-Main-Sequence (He-ZAMS) masses and metallicities. The range of initial masses comprises 20, 40, 60, 80, 100, 120, 200\Mdot\ with a comparable range of alternative models provided in table \ref{masstable}. The initial composition is based on tables provided by \cite{Grev98} with \Zdot\ $=$ 0.014, and Y $=$ 0.266. We also employed OPAL opacity tables from \cite{RogersNayfonov02}. We calculate models for 5 solar-scaled metallicities: \Zdot, 0.5\Zdot\ (LMC), 0.2\Zdot\ (SMC), 0.1\Zdot, and 0.02\Zdot, providing a range of low metallicity environments which may host the heaviest stellar mass black holes. 

We adopt the standard mixing-length theory of convection by \cite{BohmVitense58} with $\alpha_{\rm{MLT}}$ $=$ 1.5, implementing the Ledoux criterion, allowing for semiconvective mixing which we include here with an efficiency parameter of $\alpha_{\rm{sc}}$ $=$ 100, favouring blueward evolution \citep[see][]{HigginsVink2020, Schoot19}. We consider additional mixing by convective core overshooting on the main sequence, but do not include overshooting of the Helium core.

\subsection{Pre-Helium evolution}
To create a grid of He star models, we evolve H-burning models towards WR stars via extreme mixing, which promotes bluewards evolution due to additional H dredged into the core. Rather than inducing rapid rotation, we employ an artificially large increase of the convective core by exponential overshooting in our Standard Grid of models presented in this work. We include core convective overshooting above the H-burning core with a  diffusive method for values of $f_{\rm{ov}}$ up to 0.9. In principle this extreme mixing could be achieved in Nature by various paths, including strong winds, rapid rotation, and/or binary evolution. In order to aid evolution towards a He star stage without numerical complexities in the expanding stellar envelope we included the option of MLT$^{++}$ in MESA which increases the transport of convective energy in low-density envelopes.

Rotation was included with angular momentum transport and chemical mixing coefficients from \cite{Heger00}. In our Standard Grid of He star models, for all masses and metallicities the initial rotation rate is set to 20\% critical, whereas for our Alternative Grid (see Sect. \ref{QCHEgrid}) of quasi-chemical homogeneously evolved (QCHE) models we implement higher values of 60\%, with 40\% at the highest masses $>$120\Mdot. While increased mixing by rotation promotes evolution towards the He main sequence, the core He-burning models have sufficiently spun down due to angular momentum loss by main sequence stellar winds such that the varied initial rotation rates are all reduced to $\leq$ 150\kms.

For our Standard Grid of He stars, the initial H-burning ZAMS models implement zero mass loss in order to create He star models which remain massive enough on the He ZAMS to explore the full range of initial masses at all Z. The standard mass-loss recipe enabled as part of the 'Dutch' wind routine by default in MESA \citep[e.g.][]{Pax15} would implement \cite{Vink01} rates during H-burning, followed by that of \cite{NugisLamers} which sets the mass-loss rate for hot stars with less than 40\% surface H, encompassing the WR regime. Note that the default implementation of the \cite{NugisLamers} recipe in MESA (used in this study) includes a Z-dependency in terms of the current metallicity. This leads to almost no Z-scaling for the NL00 mass-loss rates due to self-enrichment. Other studies \citep[e.g.][]{Yusof13} consider the initial metallicity for the Z-dependency of \cite{NugisLamers} \citep[following][]{EV2006}, in line with the $Z_{\rm{Fe}}$-dependence from \cite{Vink05} which leads to a stronger Z-dependence of the \cite{NugisLamers} rates.

\begin{figure}
    \centering
    \includegraphics[width=\columnwidth]{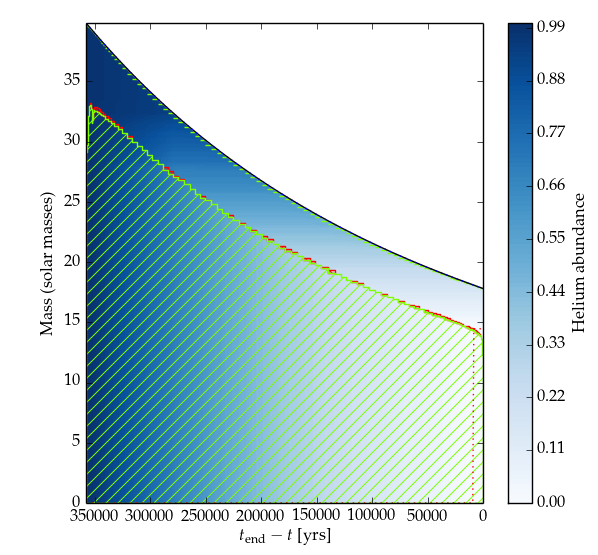}
    \caption{Structure evolution (Kippenhahn) diagram of a 40\Mdot Helium star at \Zdot with mass-loss rates from \protect\cite{SV2020}. The colour bar denotes the Helium abundance and the green hatched region illustrates the convective core size.}
    \label{newKipp}
\end{figure}

In this study, we do not discuss any explicit impact of different wind mass loss treatments before the He ZAMS, but instead focus on the impact of stellar winds after the stars have become He stars (cf. Sect. \ref{MasslossSect}). Figure \ref{newKipp} illustrates the structure evolution of a 40\Mdot\ helium star at \Zdot. The green hatched lines represent the convective core, with the helium abundance shown in blue by the colour bar. The model shown in Fig. \ref{newKipp} implements mass loss from \cite{SV2020}, and gives a final mass of 17.8\Mdot\ at core He-exhaustion. 

\subsection{Helium star evolution}
In all cases our models evolve from core H exhaustion towards the He ZAMS position at log T$_{\rm{eff}}$ $\sim$ 5.0 and continue their evolution as a Helium star until core He exhaustion where the evolution is stopped. Our models do not evolve to cooler T$_{\rm{eff}}$ towards RSG evolution after core H-burning given that the scope of this study is to test He-rich WR stars from the He ZAMS to He exhaustion. As outlined above, we do not initiate our models on the He ZAMS. Artificially altering the Helium abundance does not account for the by-products created during core H-burning ($^{14}$N) and would not provide a comparable structure to evolved WR stars. Therefore, we provide two comparable methods of producing pure He stars from the H ZAMS to the He ZAMS.

We present our Standard Grid of models in Sect. \ref{newgridSect} which omits stellar winds during the core H-burning stage of evolution in order to probe the entire mass range for Helium stars on the He ZAMS at all Z. We provide an Alternative Grid in Sect. \ref{QCHEgrid} which does include stellar winds during core H-burning, highlighting an Alternative method of creating He star models through the addition of rapid rotation. For both grids of models we focus on the core He-burning stage of evolution where we test three wind recipes for WR stars. As a result, we do not probe the evolution during core H-burning but rather compare the differing methods of reaching the He ZAMS. We find that the conclusions are interchangeable, irrespective of which method of fully mixing H-burning stars towards pure He stars is implemented. The benefit of our Standard Grid is the ability to test the evolution of 50-200\Mdot\ He stars at \Zdot, whereas the Alternative Grid approach limits the derived He star mass range due to the inclusion of mass loss, providing insights into the possible realistic initial masses of He stars. This is severe for high metallicities and would limit our studies to He ZAMS masses of about $\sim$20 \Mdot\ at \Zdot.

\section{Wolf-Rayet mass-loss rates}\label{MasslossSect}

In contrast to OB-type stars, the theory for classical WR winds is still in its infancy \cite[e.g.][]{NL2002,GH2005,Graef2017,Grassitelli+2018,Sander+2020}. Consequently, empirically determined mass-loss recipes are used in stellar evolution models. The recently derived formulae for massive He stars winds by \cite{SV2020} mark the first theoretically rooted recipes in this field. In this study, we present a comparison of the most utilised recipes for WR mass loss alongside the new prescription from \cite{SV2020}.

Based on empirical mass-loss rates obtained from radio fluxes with a non-trivial clumping correction \citep{Nugis+1998} and emission lines, \cite{NugisLamers} derived relations for the mass-loss rates of WN and WC stars including a combined recipe of the form $\dot{M}(L,Y,Z)$. The latter is commonly applied in stellar evolution codes including MESA \citep{Pax13}, GENEC \citep{Ekstroem+2012}, FRANEC \citep{Chieffi}, STAREVOL \citep{Martins17}, or PARSEC \citep{Chen+2015}. 

Another widely-used recipe, commonly applied in population synthesis modelling \citep[e.g.][]{Belczynski+2010}, is a combination of a clumping-corrected empirical mass-loss relation of \cite{Hamann95} with the iron-metallicity ($Z_{\rm Fe}$-dependent mass-loss scaling of \cite{Vink05}. Based on the tailored atmosphere analysis of a WN sample, \cite{Hamann95} derived a recipe of the form $\dot{M}(L)$, which can be extended to the form $\dot{M}(X, L)$ when accounting for a fit of the hydrogen content with the mass loss. Although various later studies with more elaborated models were made \citep[e.g.][]{Crowther+2002,Sander14,Hainich15}, it is -- with certain adjustments discussed below -- still employed in most population synthesis models.
Given the fact that the work of \citet[][]{Hamann95} was performed before the introduction of clumping and iron-line blanketing in model atmospheres -- two major additions that considerably affected the derived stellar and wind parameters -- the original values were actually overestimating the mass loss of the WR stars. Moreover, \citep{Vink05} obtained a clear metallicity-dependence of WR winds, which has to be taken into account. At solar metallicity, \citet{Yoon05} showed that scaling down the results from \citet[][]{Hamann95} by a factor of $10$ would align them more with the recipe form \citet{NugisLamers}. This scaling was then combined in \citet{Yoon06} with the $Z_\mathrm{Fe}$-scaling from \citet{Vink05}. 

With these mentioned adjustments, both widely used recipes essentially give values of $\dot{M}$ that are in the regime of observed WR stars in the Milky Way. However, this does not imply that their scalings are necessarily physically meaningful. When employing current stellar atmosphere models with prescribed wind velocity fields, there is considerable scatter in the obtained results \citep[e.g.][]{Tramper15,Hamann+2019,Shenar+2019} with regressions yielding very different $L$-dependencies from those in \cite{NugisLamers}. Mitigating these was the effort of \cite{Tramper16}, though the introduced dependencies are still empirically and not physically motivated, thus making extrapolations beyond the regime of observed objects highly questionable \citep{Vink2017}. 

More recently, \cite{Yoon17} compared the mass-loss rates of various WR types (e.g. WNE, WC, WO) derived from stellar atmosphere calculations \citep[see][]{Hamann06, Sander12, Hainich14, Hainich15, Tramper15} with theoretical wind recipes as implemented in stellar evolution codes. \cite{Yoon17} explored a range of metallicity, surface helium and luminosity dependencies on WR mass-loss rates, correlated with populations of WNE stars and WC/WO stars from stellar atmosphere models, and the implications for populations of SNe progenitors. Finding a steep dependence on initial metallicity of WNE stars compared to WC/WO stars, \cite{Yoon17} combined the relation of \cite{Hainich14} and \cite{Tramper16} with the earlier results from \cite{NugisLamers} and an additional scaling factor motivated by the lower clumping used in \cite{Hamann06}. Beside this additional scaling, the resulting recipe essentially switched from \cite{NugisLamers} to  \cite{Tramper16} at $Y = 0.9$, thereby resulting in higher mass-loss rates for WC/WO stars.

The recipe suggested by \cite{Yoon17} was created to yield evolutionary tracks surpassing the observed WR luminosity ranges in the Milky Way and the LMC -- in particular when making use of a scaling factor larger than unity. As such, it is neither directly based on empirically determined wind parameters, nor on underlying theoretical considerations. Its parameter dependence results from the underlying incorporated recipes, making it not well suited for extrapolations. The recipe by \cite{Yoon17} reflects the empirical finding that observed WC stars have higher mass-loss rates than WN stars with the same luminosity. However, this does not imply that a WN and WC star with the same $L$ also have the same mass. In fact, \cite{Sander+2020} could show with hydrodynamically consistent atmosphere modelling, that the mass loss of WC stars is lower than the one for WN stars with the same luminosity and mass (in the same host environment). Thus, the empirical result of higher mass-loss rates for WCs compared to WNs of the same luminosity and the same initial metallicity implies that these WC stars likely have a lower current mass than WN stars. In light of the results obtained by \cite{Sander+2020}, the increase of $\dot{M}$ at the onset of the WC stage in a mass-loss recipe is questionable. In this effect, the \cite{Yoon17} recipe is similar to \citet{NugisLamers}, despite the difference in individual coefficients. We thus refrain from testing a larger range of recipes in this work and only compare the new theoretical relation with the two most widely used recipes outlined in the following Sect. \ref{3recipes}.

\begin{figure}
    \centering
    \includegraphics[width=\columnwidth]{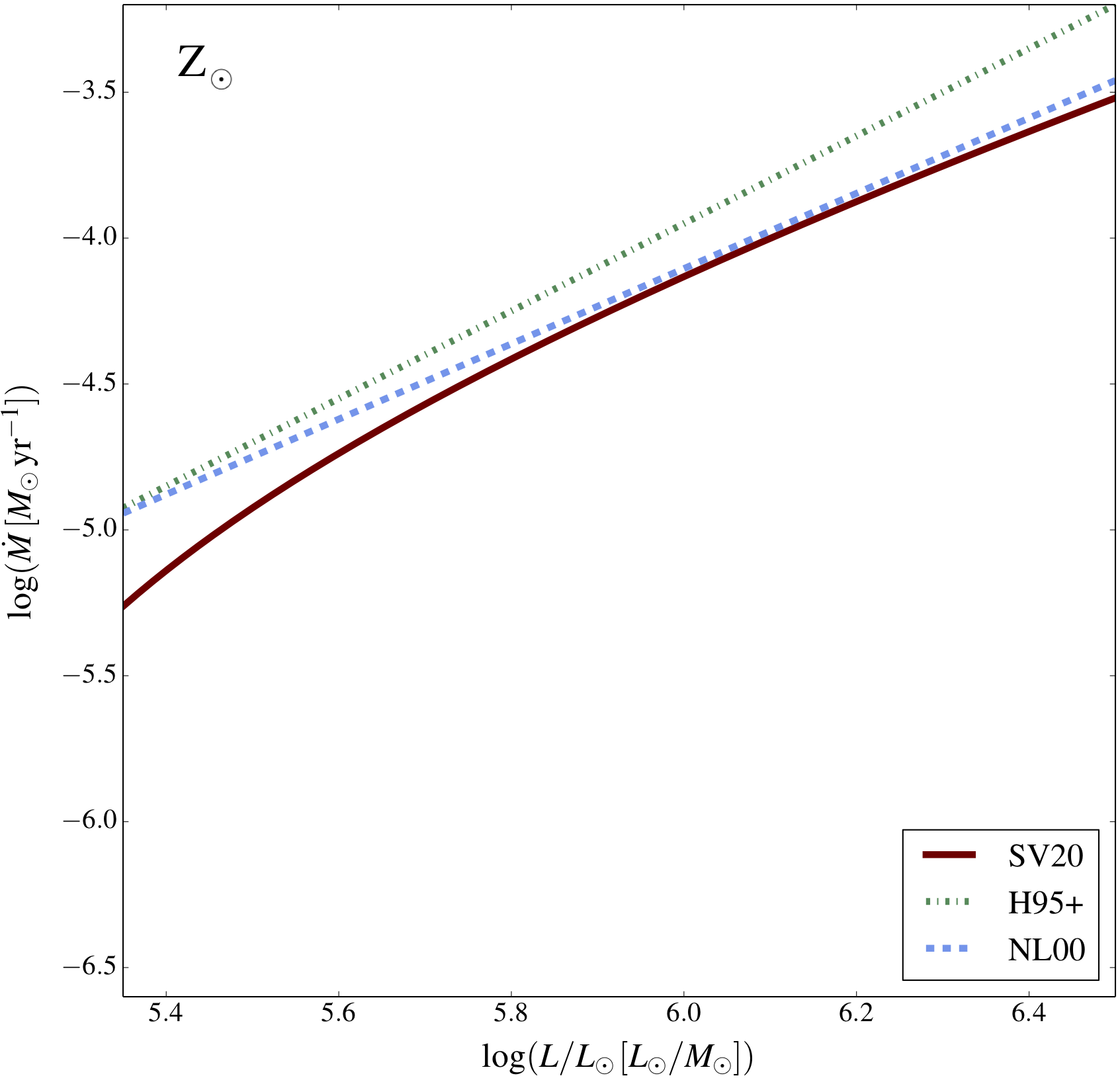}
    \caption{Mass-loss rate as a function of luminosity at the start of the He ZAMS with \Zdot\ for each wind prescription considered in this work SV20 (red), H95+ (green), and NL00 (blue).}
    \label{MdotL_sol}
\end{figure}

\begin{figure}
    \centering
    \includegraphics[width=\columnwidth]{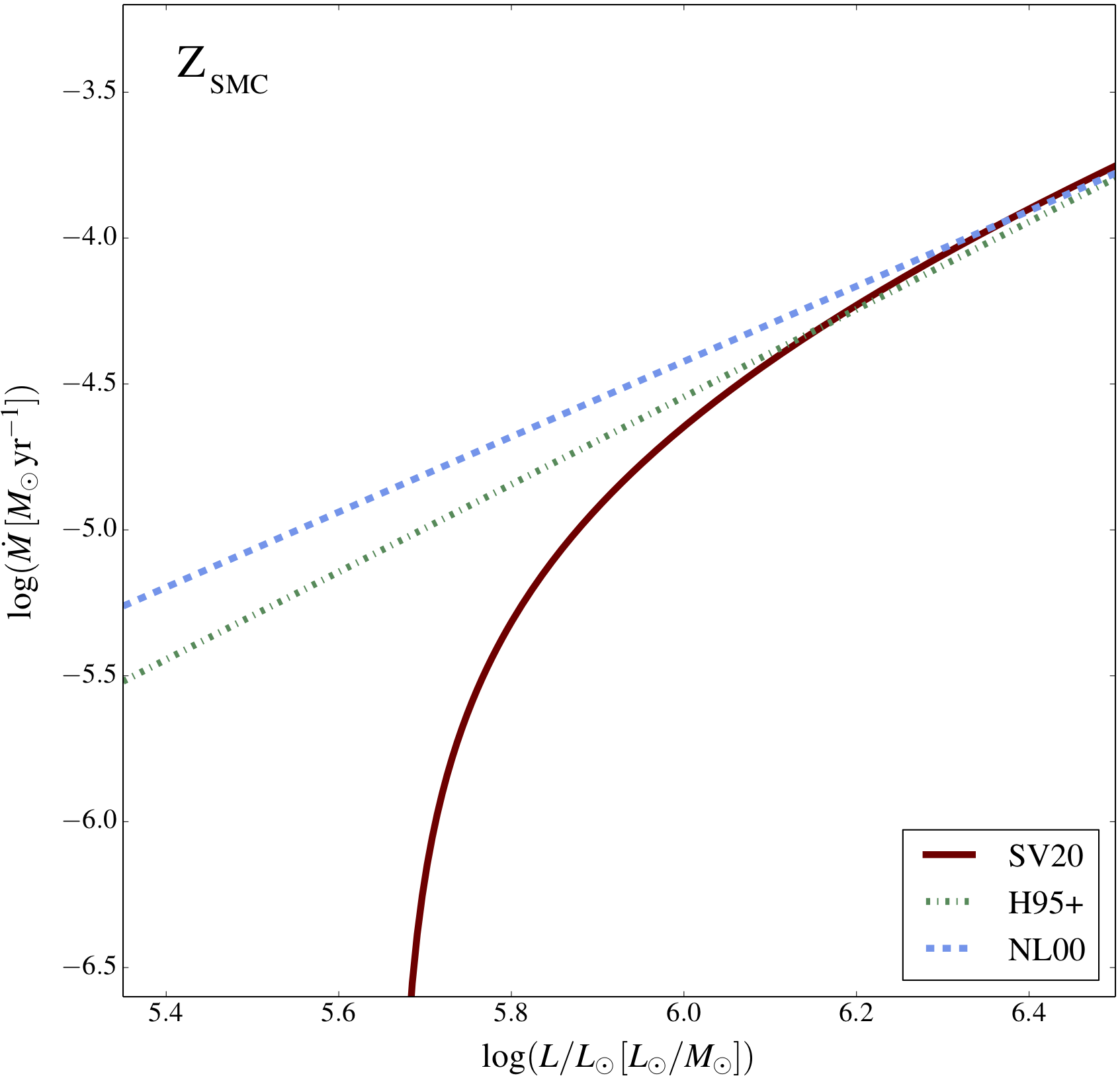}
    \caption{Mass-loss rate as a function of luminosity at the start of the He ZAMS with \ZSMC\ for each wind prescription as described in Fig. \ref{MdotL_sol}.}
    \label{MdotL_SMC}
\end{figure}

\subsection{WR mass-loss prescriptions}\label{3recipes}

In this paper, we study how much the two mainly used recipes are affecting the resulting final masses (cf. Sect. \ref{BHmassesgrid1}) of He stars and how these results change when using a physically motivated recipe based on hydrodynamically consistent stellar atmospheres. For the latter, we take the $\dot{M}(L)$-recipe provided by \cite{SV2020} (hereafter SV20).
\begin{equation}
    \label{Seq}
    \dot{M}_{\mathrm{SV20}} = \dot{M}_{10}\ \left(\log \frac{L}{L_{0}}\right)^\alpha \left(\frac{L}{10 \cdot L_{0}}\right)^{3/4} 
\end{equation}
with coefficients $\dot{M}_{10}$, $L_{0}$ and $\alpha$ provided for a range of metallicities in table \ref{coefftable}.
In this work, we are studying the imprint of mass loss on the He main sequence, where our stars essentially follow a clear $M$-$L$-relation (cf.\ Sect.\,\ref{sec:mlrel}). Thus, we can implement the mass loss recipe from \citet{SV2020} in the $\dot{M}(L,Z_\mathrm{init})$-form, i.e.\ as a function depending only on luminosity $L$ and initial metallicity $Z_\mathrm{init}$. This is further justified as stars which have lost all of their hydrogen layers will spend most of their remaining life time on the He main sequence. However, we stress that the treatment used in this work does not account for other evolutionary stages, e.g.\ stars beyond central He burning or stars which have retained part of their hydrogen envelope, as their $M$-$L$ relation can be significantly different and the change of their temperature and chemical abundances would have to be taken into account. 
In this work, we use a dedicated set of input parameters given in Table \ref{coefftable} for the free parameters in Eq.\,(\ref{Seq}). We apply these values in a MESA subroutine where we switch on this mass-loss recipe during the core He-burning phase. There have been minor updates in SV20 to the specified coefficients outlined in Table \ref{coefftable} for the finalised set of atmosphere models presented in SV20.
However, we tested multiple calculations to verify that these minor differences are indistinguishable in their model results. For future applications, we suggest to use the metallicity fit relations provided in SV20 which are also implemented in the publicly available \texttt{Python} script.\footnote{The script is available at \url{https://armagh.space/asander}}

In MESA, the standard mass-loss recipes are accumulated in the so-called `Dutch' wind scheme, which  employs \cite{Vink01} for hot and H-rich stars, \cite{deJager} for cool stars, and the aforementioned \cite{NugisLamers} (hereafter NL00) for hot, H-depleted stars. NL00's implementation in MESA has the form
\begin{equation}
  \label{NLeq}
  \log\dot{M}_\mathrm{NL00} = -11.00 + 1.29\ \log\ L + 1.73\ \log\ Y + 0.47\ \log\ Z_\mathrm{cur}
\end{equation}

where it should be noted that $Z_\mathrm{cur}$ here refers to the {\it total} current $Z$, i.e. it is enhanced due to self-enrichment of elements such as carbon, which is no longer considered to be the physically dominant component \citep{Vink05}. 

The final set of comparison models implements mass-loss rates as suggested by \cite{Yoon06} where the recipe from \cite{Hamann95} has been reduced by a factor of 10 in order to account for clumping. The resulting wind recipe implemented in stellar population models, often simply referred to as `Hamann 95' (hereafter H95+, where the '+' represents the adjustments for clumping and Fe-scaling), thus reads
\begin{equation}
  \label{Heq}
  \log\dot{M}_\mathrm{H95+} = -12.95 + 1.5\ \log\ L - 2.85\ X + 0.85\ \log\ (Z_\mathrm{init}/\Zdot)
\end{equation}

\begin{table}\caption{Coefficients at various $Z$ for $\dot{M}$(L) by SV20. (Minor revisions have been made to the coefficients in the resulting formulae of \protect\cite{SV2020}, however while we have tested the updated relations, the results are indistinguishable.)}
    \centering
    \begin{tabular}{c|c|c|c}
        \hline
         $Z_\mathrm{init}/$\Zdot & $\alpha$ & $\log L_{0}$ & $\log \dot{M}_{10}$ \\
         \hline
         1.0 & 1.301 & 5.043 & -4.075 \\
         0.5 & 1.327 & 5.335 & -3.812 \\
         0.2 & 1.299 & 5.668 & -3.523\\
         0.1 & 1.165 & 5.906 & -3.331 \\
         0.02 & 0.938 & 6.478 & -2.941 \\
         \hline
    \end{tabular}
    \label{coefftable}
\end{table}

In summary, we evolve stellar models toward the He-ZAMS and probe the core He-burning evolution with three wind recipes:

\begin{enumerate}
    \item SV20: Physically motivated mass-loss description resulting from detailed sets of dynamically consistent atmosphere models by \cite{SV2020}.
    \item NL00: Standard recipe implemented in stellar evolution codes from \cite{NugisLamers}, although accounting for total $Z$ (self-enriched) rather than initial Fe-scaled $Z_\mathrm{init}$.
    \item H95+: \cite{Yoon06} implementation as often used in Population Synthesis modelling. Rates have been extracted from \cite{Hamann95}, reduced by a factor 10 to account for clumping, and including the Fe-dependent scaling from \cite{Vink05}.
\end{enumerate}

\subsection{Comparison of wind recipes}
The key points to consider regarding the impact of different mass-loss recipes are the absolute mass-loss rates and their $Z$-dependence including in particular the actual implementation for the latter. While mass-loss recipes are written in the form of assigning a mass-loss rate to a given abundance, it is in reality the resulting opacity that drives the mass loss. Thus, depending on the ionisation stage, not every element available at the surface will actually provide significant line opacity to drive a stellar wind. The works of \cite{Vink05} and \cite{GH2005} could show that decisive line opacities in the winds of classical WR stars stem from iron and not from other elements such as CNO despite their large abundances. Performing detailed, locally consistent calculations, \cite{Sander+2020} recently demonstrated that due to the deep launching of the winds, the ionization stage of many elements at the sonic point is simply too high to provide considerable line opacities, while the iron-group M-shell ions can still contribute to the total line opacity there. While the CNO abundance is increased due to self-enrichment, the abundance of the iron group elements are confined to their initial value, at least when neglecting external influences such as stellar mergers or the pollution due to the infall of a planetary companion. Consequently, the $Z$-scalings derived by \cite{Vink05} and \cite{SV2020} should be performed with the \emph{initial} $Z$ (or simply $Z_\mathrm{Fe}/Z_{\mathrm{Fe},\odot}$, i.e. the iron mass fraction relative to the solar value), not the current surface abundances. This is not the case in the recipe by \cite{NugisLamers} and while it was adopted by \cite{Yoon06}, this greater insight has not been universally adopted in the literature.

In the recent mass-loss recipe by \cite{SV2020}, the $Z$ (i.e. $Z_\text{init}$) dependence is physical, and shown to be more complex than that of a simple power law. This is illustrated in Figs.\,\ref{MdotL_sol} and \ref{MdotL_SMC} where we plot the $\dot{M}(L)$-behaviour of the different recipes at the He ZAMS for solar and SMC metallicity. While both NL00 as well as H95+ use power-law scalings, the SV20 curves reflect the finding of \cite{SV2020}, highlighting when a star has a sufficient $L/M$-ratio to reach WR-type winds, which is particularly noticeable in Fig.\,\ref{MdotL_SMC} depicting the SMC-like case. However, once the regime of dense WR winds is fully reached, the $Z$-dependence in SV20 is flatter than any other current $\dot{M}$-recipe with its mass loss changing only by a factor of two for an order of magnitude change in metallicity.

Focusing on the resulting range of BH masses, \citet[][]{Woosley+2020} investigated the effect of He star mass-loss when employing different recipes. While they did not calculate detailed structure models for each mass-loss recipe, the overall loss of mass during the He-burning stage could be approximated sufficiently by using previous work \citep[][]{Woosley19} yielding the luminosity as a function of mass and lifetime. The initial findings from \citet[][]{Sander+2020} -- in particular the strong deviation from a power-law in $\dot{M}(L/M)$ -- were already taken into account in \citet[][]{Woosley+2020}, but the underlying model sequence was small and the models were not tailored to represent ($L$,$M$)-combinations along He ZAMS as later done in \citep[][]{SV2020}. We avoid these obstacles \citep[e.g. the limited ability to extrapolate the results from ][]{Sander+2020} in our current work by employing the newer calculations from SV20. The imprint on the resulting BH masses between \citet[][]{Woosley+2020} and our work can be quite large, in particular for higher masses, where SV20 predict much higher mass-loss rates than what would be inferred from extrapolating the \citet{Sander+2020} relation. This discrepancy is rooted in the underlying polynomial fit for $\dot{M}$ in \citet{Sander+2020}, which eventually yields a local maximum in $\dot{M}$. Higher mass calculations in SV20 could instead show that such a maximum does not exist and a different mathematical description had to be employed for $\dot{M}$ to avoid such an incorrect `asymptotic behaviour'.

\subsection{The Mass-Luminosity relation}\label{sec:mlrel}
In this study, we provide stellar evolution models with mass-loss rates based on \cite{SV2020}. The mass-loss recipe in \cite{SV2020} was derived from a set of atmosphere models assuming an $M$-$L$-relation for chemically homogeneous helium stars by \cite{Graefener+2011}. To test whether this $M$-$L$-relation is actually representative for our structure models on the He ZAMS, we compare our results of ($L$, $M$)-tuples with the relation from \cite{Graefener+2011} in the Appendix (see Fig.\,\ref{GotzML}). The comparison highlights an excellent overall agreement between the M-L relation for He stars with our models (blue triangles) and that of \cite{Graefener+2011} (black solid line) for a broad range of masses and metallicities.

\begin{figure*}
    \centering
    \includegraphics[width=17cm]{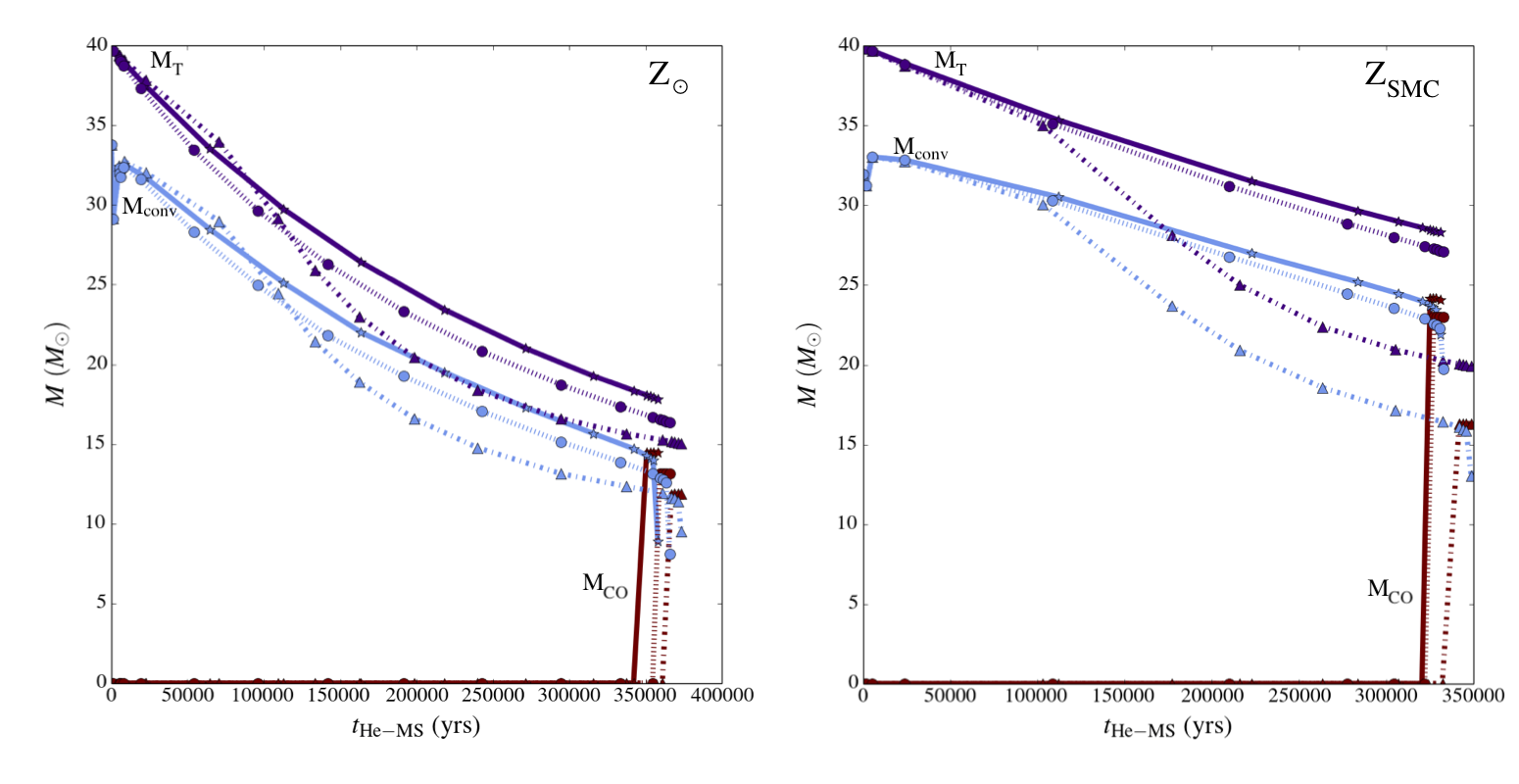}
    \caption{Structure evolution of a 40\Mdot\ star with each mass-loss recipe (solid line with starred markers represent SV20, dashed dotted lines with triangle markers represent NL00, and dashed lines with circle markers employ H95+ rates). Left: models with \Zdot. Right: \ZSMC\ models. The purple lines show the total mass (M$_{\rm{T}}$), blue lines outline the convective core mass (M$_{\rm{conv}}$), and red lines show the final CO core mass (M$_{\rm{CO}}$) at He exhaustion. The x-axis timescale illustrates the evolution from the He-ZAMS until core He-exhaustion. }
    \label{Kipp40}
\end{figure*}

\begin{figure*}
    \centering
    \includegraphics[width=17cm]{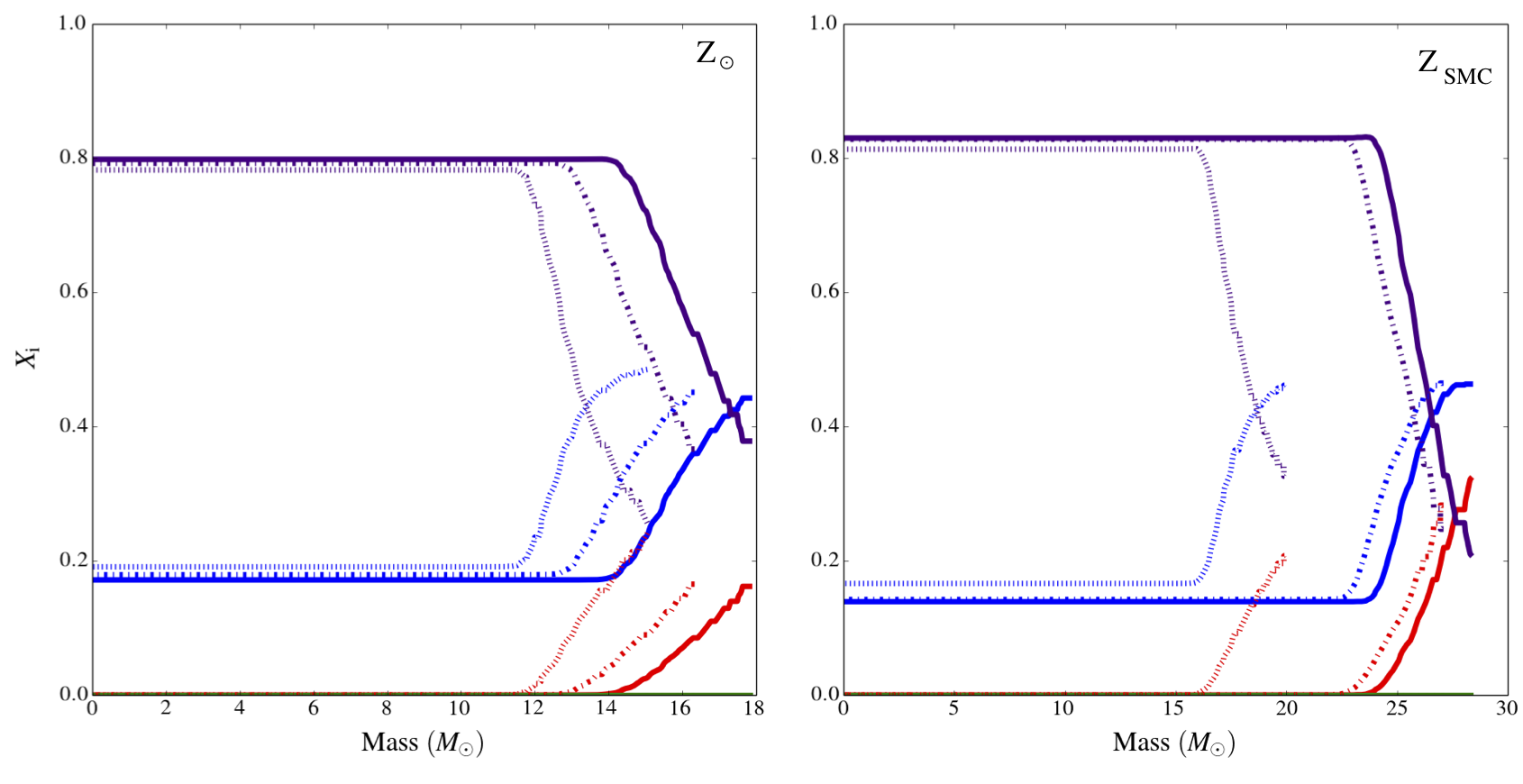}
    \caption{Final abundance profile of a 40\Mdot\ star at core He-exhaustion with each mass-loss recipe (solid lines represent SV20, dashed lines represent NL00, and dashed-dotted lines show H95+ models). Left: models with \Zdot. Right: \ZSMC\ models. The red lines show $^{4}$He abundances, blue lines illustrate $^{12}$C, and purple lines show $^{16}$O. The x-axis illustrates the mass co-ordinate from core to surface at core He-exhaustion. }
    \label{Abundance40}
\end{figure*}

\section{Results from the Standard Grid}\label{newgridSect}

For our Standard Grid of helium WR stars, we have calculated a wide range of models with initial masses 20-200\Mdot, at \Zdot, 0.5\Zdot, 0.2\Zdot, 0.1\Zdot, 0.02\Zdot. We evolve H-burning models towards becoming He stars via extreme artificial convective mixing as outlined in Sect. \ref{SectModels}. These He-ZAMS models have less than 1\% surface H at all $Z$ and for all initial masses by the He-ZAMS. We now compare the evolution during the He-main sequence, and ultimately the final masses which may be utilised as a proxy for BH progenitor masses since the final evolutionary stages post$-$He burning are extremely short.

The final mass at core He-exhaustion changes drastically with decreasing $Z$ for the different wind recipes. At \Zdot, the highest final masses are provided by models with SV20 winds, with models implementing NL00 showing the lowest final masses. The situation is comparable at SMC-like Z, where SV20 models show the highest final masses, closely followed by those with H95+ winds, but NL00 models remain the lowest mass models at He-exhaustion. At the lowest Z, we find that H95+ models have higher final masses than models which include SV20 rates. This is due to the steeper L-dependencies of H95+, compared to the theoretical recipe by SV20.

\subsection{Structure and abundances}
We compare the structure evolution of a 40\Mdot\ model, as in Fig. \ref{newKipp} for SV20, with the subsequent two recipes in Fig. \ref{Kipp40}, for \Zdot\ (left) and \ZSMC\ (right). This figure highlights the change in final mass and core mass with $Z$ and mass-loss rate. The total mass is shown in purple with the solid lines representing SV20 models, dashed-dotted lines illustrating models with NL00, and dashed lines showing H95+ models. Convective core masses are also provided in blue, with final CO core masses shown in red. We show here that at \Zdot\ (left), the final mass is relatively similar for all recipes, with a variation of 2-3\Mdot. However, the situation shifts at lower Z, where even at \ZSMC\ (right) the change in final mass can be 8-10\Mdot. The drop in final mass can be seen for the NL00 model (dashed dotted) compared with those of SV20 and H95+ which have a much more shallow decline in total mass during core He-burning.

Another interesting consequence of the different mass-loss treatments is the resulting abundance profile at He-exhaustion. Figure \ref{Abundance40} shows the chemical abundance profile as a function of the mass coordinate throughout the star (from core to surface) for the same models as in Fig.\,\ref{Kipp40}. $^{4}$He is represented by red lines, $^{12}$C is shown by blue lines, and $^{16}$O in purple. Once again, the clear trend in mass loss between the three different recipes is clearly visible. In particular, one can notice the strong stripping of outer layers with the NL00 formula even at \ZSMC\ (right panel). Both at \Zdot\ as well as at \ZSMC\,  the dominant surface abundance at He exhaustion is carbon with the resulting value being almost independent of the mass-loss recipe and metallicity. However, this finding does not hold for other mass ranges. For our $60\,M_\odot$ models, the stripping is stronger and oxygen can become the most abundant element at the surface. On the contrary, the stripping is weaker for the $20\,M_\odot$ case, where WR-type mass-loss even breaks down completely at \ZSMC\ and we are left with a He-dominated atmosphere at all metallicities except \Zdot. 

\begin{figure}
    \centering
    \includegraphics[width=\columnwidth]{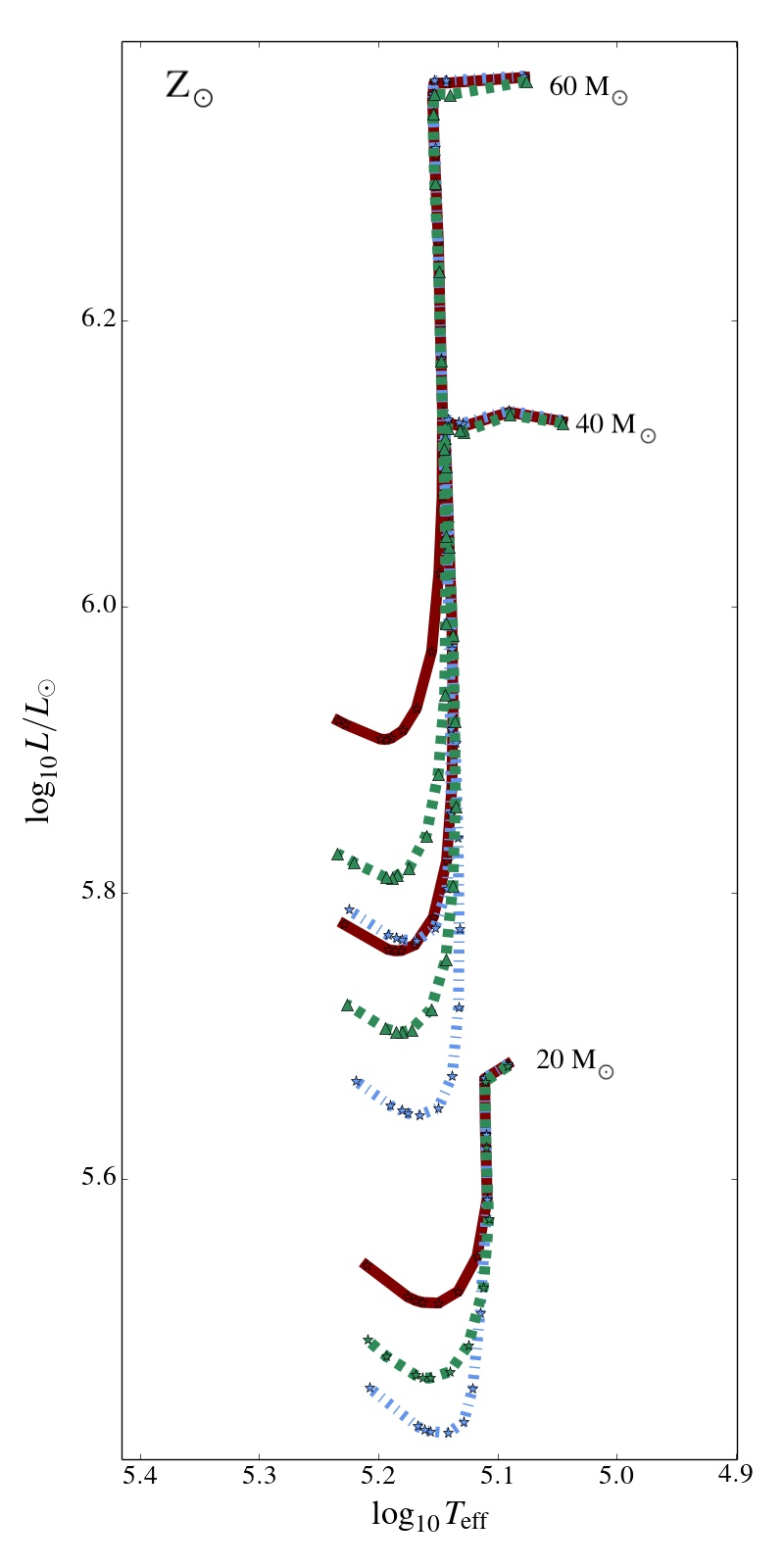}
    \caption{Hertzsprung-Russell diagram of 20-60\Mdot\ He-burning stars evolving from the He-ZAMS as Wolf-Rayets, calculated at \Zdot. Red solid tracks represent models which employ SV20 mass loss, those with NL00 rates are shown in blue dash-dotted lines, and green dashed lines illustrate models which include H95+ mass loss. }
    \label{HRDcomparison1}
\end{figure}

\begin{figure}
    \centering
    \includegraphics[width=\columnwidth]{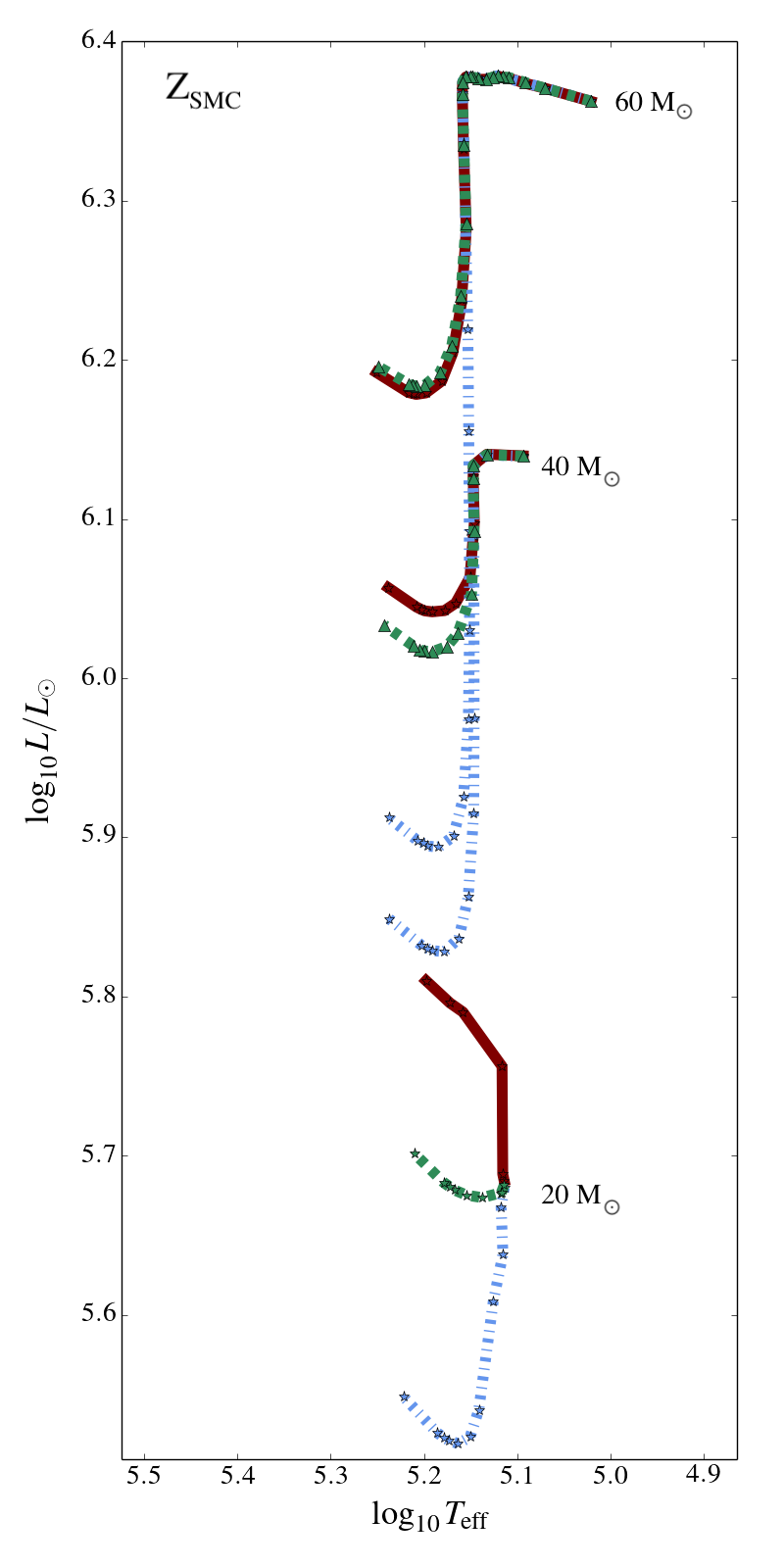}
    \caption{Same as Fig.\,\ref{HRDcomparison1}, but for models calculated at \ZSMC. At this metallicity, the SV20 description predicts no strong wind mass loss for the $20\,M_{\odot}$ model, thus yielding a completely different behaviour in the HRD.}
    \label{HRDcomparison2}
\end{figure}

Until core-He exhaustion, we never see a maximum of $X_\text{C} \approx 0.5$ at the surface as a re-occurring pattern in all of our abundance profiles. When progressing inside, the carbon fraction is getting larger while the He fraction is shrinking. After $X_\text{C}$ reaches a maximum, it declines further inwards and oxygen becomes the dominant element with relative mass fractions of around $0.8$ with only minor spreads due to mass and mass-loss treatment. Consequently, we would not expect to observe WC and WO stars with $X_\text{C} > 0.5$ at their surfaces. In their empirical analyses of WC and WO stars, both \citet{Sander12} and \citet{Tramper15} get fractions of up to $0.62$. This might very well be within the uncertainties of the present studies, but a more in-depth look in future studies with both structure and evolution models could yield important constraints on the parameters and precise evolutionary stage of known WC and WO stars. 

For our $40\,M_\odot$ models, the second most abundant element at the surface at \Zdot\ is oxygen  with mass fractions between 25\% and $\sim 40\%$. From both the temperatures as well as the surface abundance, we expect such stars to have a WO-type spectrum. Neglecting the solution resulting from the overestimation of mass loss by NL00, the situation at \ZSMC\ is quite different: The oxygen abundance is lower here and potentially even below the remaining amount of helium at the surface. While these fractions could probably still result in a WO-type spectrum, the different O/He-ratios could potentially provide an interesting albeit indirect metallicity constraint for observed WO stars. The current observational constraints \citep{Tramper15,Shenar+2016} do not contradict this scenario, but they also do not provide clear evidence. This is not surprising as the O/He-ratio at the surface also depends on the mass regime with higher mass models also showing larger oxygen fractions at the surface. Thus, a certain ratio could either be reached by a more massive star at lower metallicity or a less massive star at higher metallicity. Still, if additional constraints are available which could break this degeneracy, the surface abundances could potentially provide important indirect diagnostics to otherwise inaccessible pieces of information.

The overall evolution of our model sample in the HRD is highlighted in Figs.\,\ref{HRDcomparison1} and \ref{HRDcomparison2} where we show representative models for 20, 40 and 60\Mdot\ at \Zdot\ (left) and \ZSMC\ (right), with mass-loss recipes from SV20 (red solid lines), NL00 (blue dashed dotted lines), and H95+ (green dashed lines). Additional representative models for \Zdot\ and \ZSMC\ are shown in Figs. \ref{HRD2solar} and \ref{HRD2smc} with initial masses of 40, 80 and 120\Mdot. The drop in luminosity seen in the blue tracks (NL00 mass loss) reflects the stronger winds applied in this recipe compared to the red and green tracks. At low $Z$, models which implement the new SV20 mass loss may reach the `breakdown regime' at lower masses as the luminosity there is not sufficient to support the optically thick winds. This is evident in the qualitatively different shape of the 20\,$M_\odot$-track at \ZSMC. Without an optically thick wind, the model loses less than a solar mass throughout the whole He-burning lifetime. Consequentially, the star hardly changes its position in the HRD until the core contracts and this leads to an increase in the surface luminosity.

\begin{figure*}
    \centering
    \includegraphics[width=\textwidth]{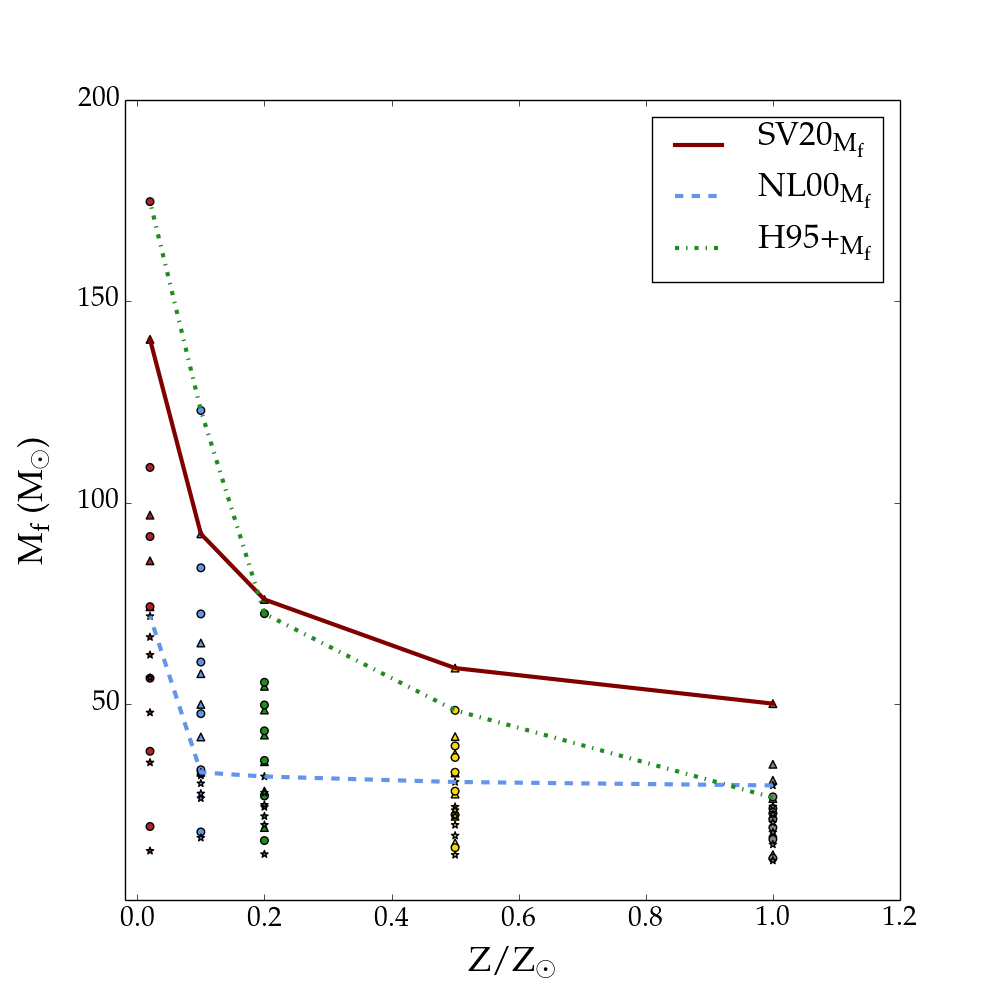}
    \caption{Final mass at core He-exhaustion as a function of $Z$ for each mass-loss recipe. The most massive final masses for each recipe are shown by the solid line in red for SV20, NL00 with a dashed blue line, and H95+ with the dash-dotted green line. Coloured markers denote the varied $Z$ for each set of models, with triangles denoting SV20 models, circles for H95+ models and finally stars for NL00 models.}
   
    \label{MfZnew}
\end{figure*}

\subsection{Final masses at core Helium exhaustion}\label{BHmassesgrid1}
The final mass at the end of core He-burning provides a much needed proxy for BH progenitor masses as the remaining nuclear burning phases are exponentially shorter than for H or He-burning. We probe the variety of final masses for our grid of WR models as a function of decreasing $Z$ to establish the upper stellar BH mass limit, before considering additional reductions due to (P)PI. By further comparing the most commonly implemented stellar wind prescriptions for hot, helium stars, we also provide a direct contrast which considers the ramifications of implementing each recipe in stellar evolution. 

In particular, when studying a wide range of initial masses across all Z, we find that the maximum final mass for models with NL00 winds is $\sim$ 30\Mdot, with the exception of the highest initial mass model at $\nicefrac{1}{50}$ $^{\rm{th}}$ \Zdot. This means that with the standard `Dutch' wind recipe implemented in MESA (also included in many other stellar evolution codes) one would need to consider very massive Helium stars (i.e. above 200\Mdot) in order to reach a final compact object mass above 30\Mdot,  or limit heavy BH progenitors to essentially primordial Z. In other words, we find that the high mass-loss rates for He stars even at low $Z$ due to the - physically incorrect - self-enrichment of the NL00 recipe has a tremendous impact on the evolutionary channels for GW progenitors as it prevents the formation of heavier black holes from He stars. In contrast, the H95+ prescription from \cite{Yoon06} provides a similar range of final masses to that of SV20. 

Figure \ref{MfZnew} depicts the range of final masses at different metallicities $Z$, highlighting the upper mass limits for each wind recipe. The solid red line highlights the maximum final mass for SV20 models at all Z, with the same shown for NL00 models in the dashed blue line, and for H95+ models in the dash-dotted green line. The coloured symbols represent the various metallicities, with triangles representing SV20 models, H95+ models with circles, and stars for models implementing NL00 rates. At \Zdot, we find that the most massive final mass is provided by SV20 models, with much lower final masses from both NL00 and H95+. However, at lower $Z$ ($\sim$ \ZSMC), the situation begins to shift towards similar final masses with H95+ and SV20 due to the different $Z$-dependence of both recipes. On the other hand, NL00 models remain below 35\Mdot\ at all $Z$ above 0.02\Zdot. For models at (and below) 0.1\Zdot, the maximum final masses are now provided by H95+, while the SV20 models remain much more massive than those of NL00.

\begin{figure}
    \centering
    \includegraphics[width=\columnwidth]{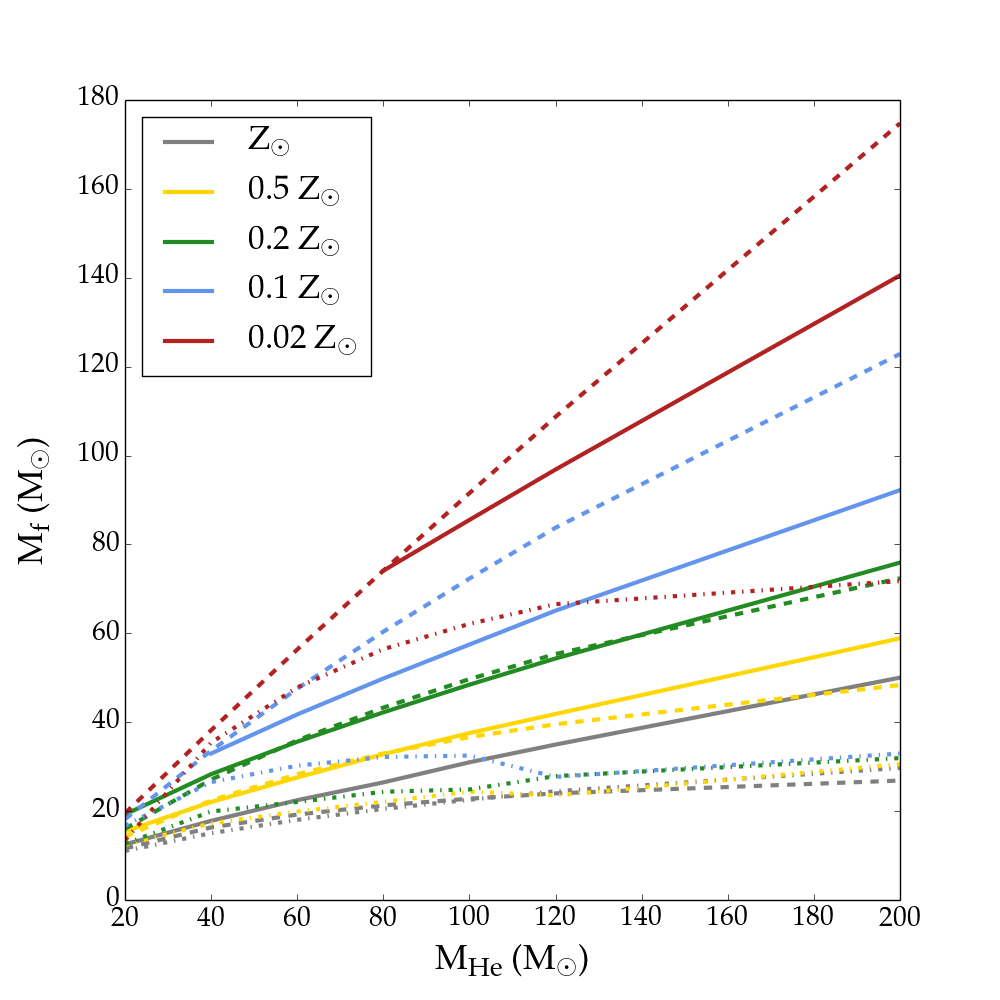}
    \caption{Final mass as a function of initial He ZAMS mass for each $Z$ and mass-loss recipe (SV20 in solid lines, H95+ in dashed lines, NL00 in dash-dotted lines).}
   
    \label{MfMhenew}
\end{figure}

Similarly, we provide the range of final masses as a function of initial He ZAMS mass in Fig. \ref{MfMhenew}. SV20 models are shown in solid lines, with NL00 in dashed-dotted lines, and H95+ in dashed lines. We illustrate that for decreasing Z, the final mass of each set of models increases due to $Z$-dependent winds. The dashed-dotted lines of NL00 do not inherently increase as in the SV20 or H95+ models due to the stronger mass loss caused by self-enrichment during core He-burning. Once again we notice that the NL00 recipe yields fundamentally different behaviour for the final masses at lower $Z$ than SV20 and H95+.

\begin{figure}
    \centering
    \includegraphics[width=\columnwidth]{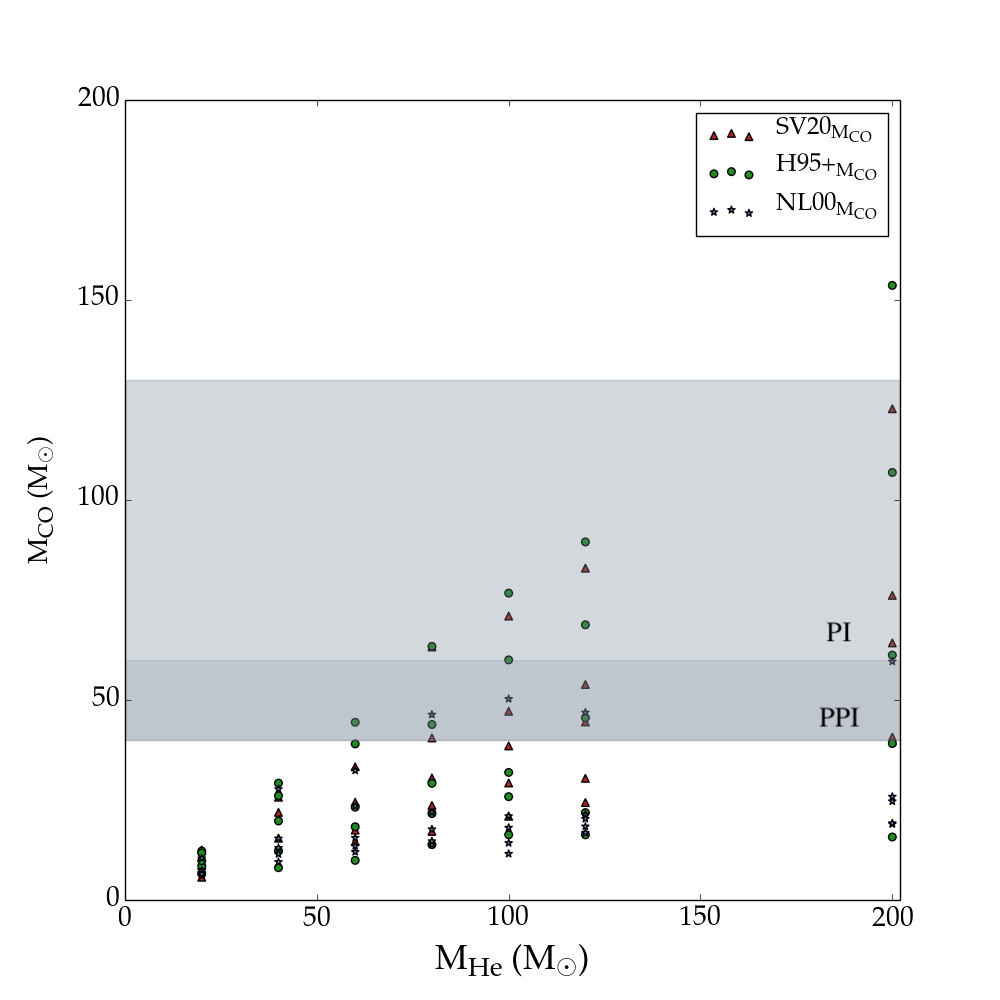}
    \caption{Final CO core mass as a function of He ZAMS mass for each mass-loss recipe (SV20 in red triangles, NL00 in blue stars, H95+ in green circles). The shaded regions correspond to mass limits where above a certain mass, larger CO core masses lead to pulsational pair instability (PPI) supernovae and pair instability (PI) supernovae respectively.}
    \label{cocore}
\end{figure}

\begin{figure}
	\includegraphics[width=\columnwidth]{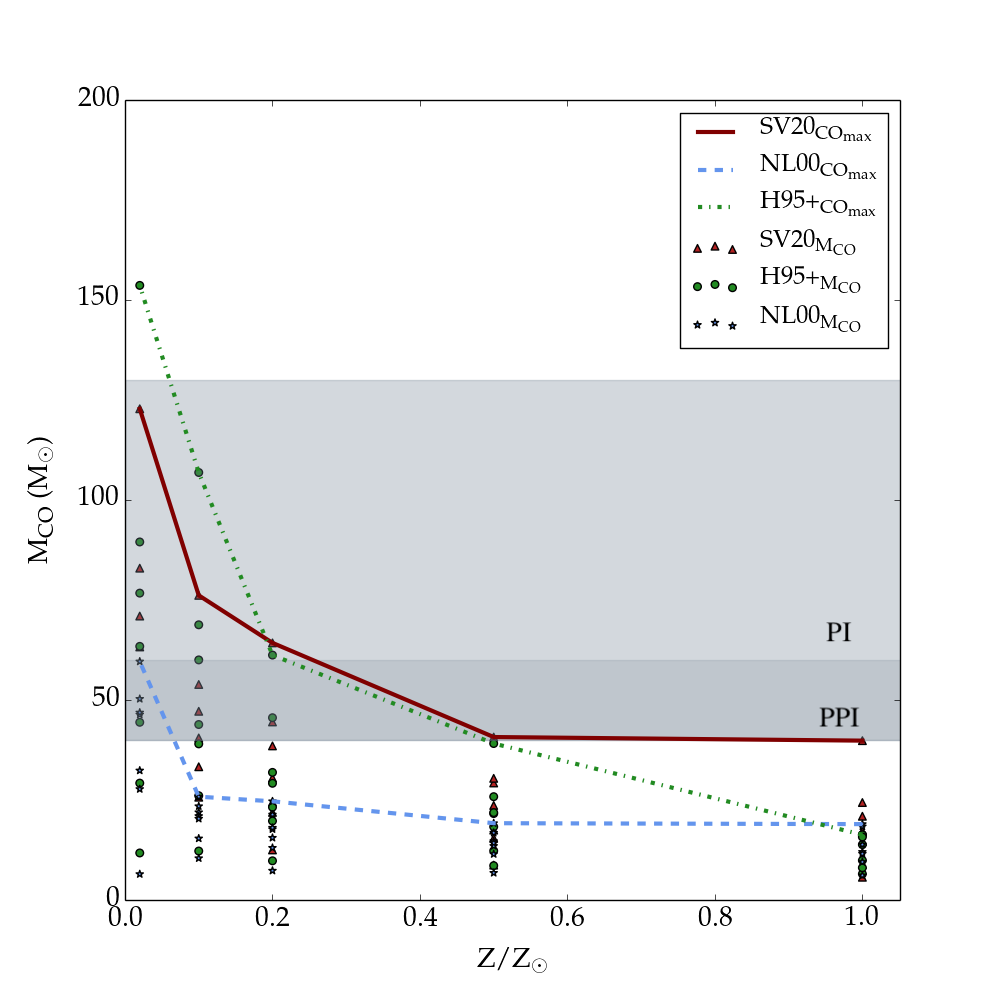}
	\caption{Final CO masses at core He-exhaustion as a function of initial metallicity for each wind recipe (SV20 in red triangles, NL00 in blue stars, H95+ in green circles). The maximum final CO core mass (CO$_{\rm{max}}$) models, for each metallicity, which include mass-loss rates from SV20 are represented by solid red lines, H95+ in dashed-dotted green lines and NL00 in dashed blue lines. The grey shaded region highlights the region where pulsational pair-instability supernovae occur, with CO core masses above 40\Mdot.}
    \label{shadedCOcore}
\end{figure}

So far, we did not take pair instability effects into account, but simply illustrated the consequences of extrapolating empirical mass-loss formulae without a deeper physical motivation. We now highlight the additional limitations due to PI in Figs. \ref{cocore}, \ref{shadedCOcore} and \ref{CloseupCOcore}, where each model's final CO core mass is shown as a function of initial He mass (SV20 shown in red, NL00 in blue and H95+ in green). These figures demonstrate that some high mass models in our grid enter the PPI regime where the CO core mass is above the estimated 40\Mdot\ limit for stripped stars \citep{Farmer+2019,Woosley2017}. We further discuss the implication for the resulting BH mass limit in Section \ref{PPI}.

\section{Results from QCHE Alternative Grid}\label{QCHEgrid} 
As an alternative approach to our standard He star grid, we have calculated another grid of models incorporating a different evolutionary path towards He stars. This `Alternative Grid' consists of rotating models with 60\% critical rotation which evolve towards becoming He stars during the MS by increased mixing. Since the critical rotation rate scales with increased mass, models which are greater than 150\Mdot\ have a reduced rotation rate of 40\% critical. Stellar winds are included during the MS, and as a result the stars spin down towards $\sim$ 150\kms\ by the onset of core He-burning such that models remain fully mixed but are no longer rapidly rotating. Our Alternative Grid mimics a channel of quasi-chemically homogeneous evolution (QCHE), which has been suggested for some Wolf-Rayet stars \citep[e.g.][]{Martins+2009,Hainich15}.

In our Alternative Grid, the mass loss during core H-burning only uses the \citep{Vink01} description and does not depend on the He star wind recipe. To ensure this, we removed the common, but physically questionable switch to a WR-wind recipe based on the hydrogen surface abundance. Since we account for main-sequence mass loss in this method, the masses of the models on the He ZAMS are lower than the initial masses of the models. Although our treatment of core H-burning can only act as a proxy for the actual processes happening in this stage, the overall loss of mass provides us with some insights about the different range of masses to be expected on the He ZAMS and the restrictions for the final masses of black hole progenitors. Although we start with initial masses up to 170\Mdot\ in our Alternative Grid, we only reach $\sim$20\Mdot\ on the He ZAMS at \Zdot\ with final masses of black hole progenitors on the order of $\sim$10\Mdot. In our stellar evolution models for 0.2\Zdot\ (SMC), the highest ZAMS masses yield a He star of $\approx 50\,$\Mdot\ with a final mass of 32\Mdot\ when applying the SV20 and H95+ wind recipes, while the NL00 recipe leads to a much lower $M_{\rm{f}}$ of $21.5$\Mdot. This effectively means a shift by about $10\,$\Mdot\ in the range of reasonably expected BH masses at SMC metallicity. The comparison between all three wind recipes alters in the lowest metallicity environments due to the $\dot{M}(Z)$ relations, with the highest resulting masses obtained in H95+ followed by SV20 and NL00, the latter two being separated by a large margin. For $2\%$\Zdot, H95+ models provide the highest masses of 120\Mdot\ by core helium exhaustion, while models with the SV20 recipe have final masses up to 100\Mdot\ for the same initial H ZAMS mass of 170\Mdot. Comparatively, models with the NL00 prescription result in a much lower M$_{\rm{f}}$ $\sim$ 28\Mdot\ leading to a change in final mass of $\Delta$ M $\approx$ 92\Mdot\ compared to that of H95+. The trend in final masses deduced from our alternative method is comparable to the results from our standard method described in Sect. \ref{newgridSect}, suggesting robust results irrespective of which method is used in creating He star models. The Alternative Grid effectively leads to a finer spacing of initial He masses at higher metallicities, giving us some additional insights. As previously seen in Sect. \ref{BHmassesgrid1} for the Standard Grid, we see that models which implement NL00 yield final masses no greater than ~30 \Mdot\ for $Z$ $\geq$ 2\% \Zdot.

\begin{figure}
	\includegraphics[width=\columnwidth]{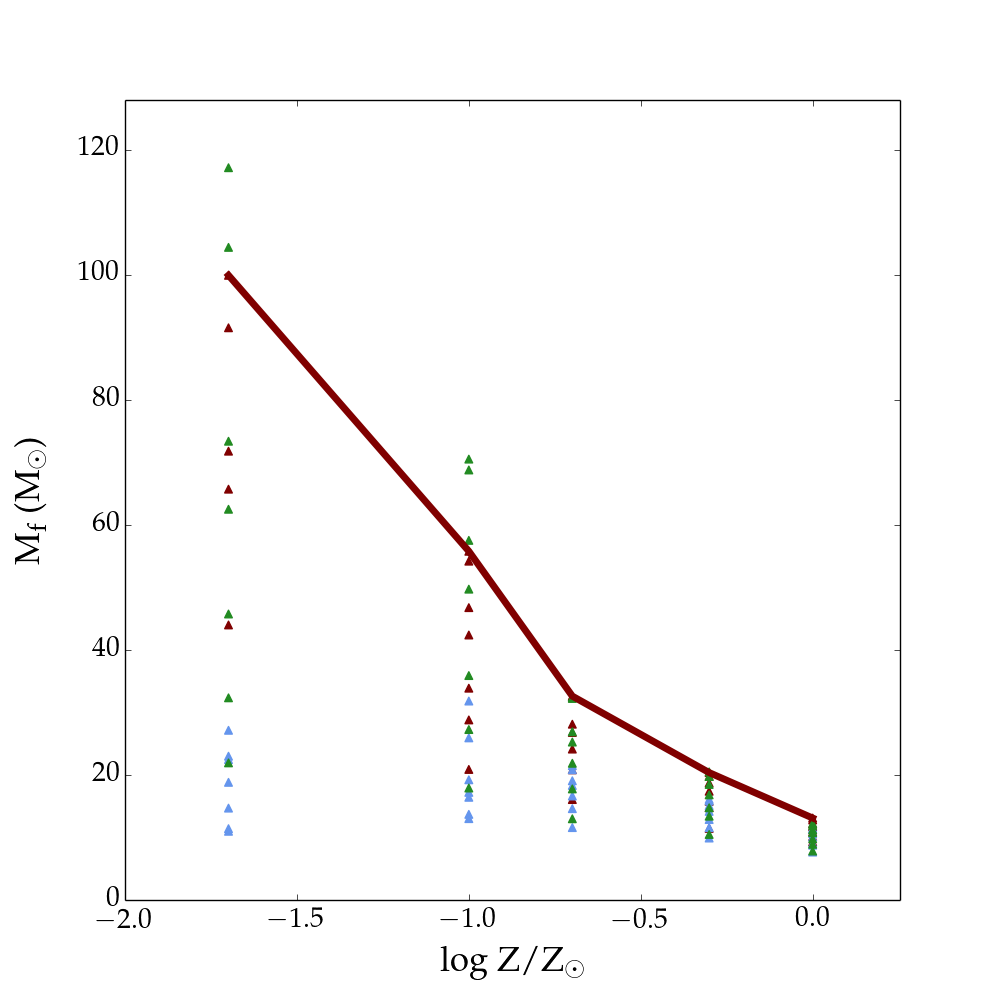}
    \caption{Maximum final mass for the Alternative Grid as a function of log Z/\Zdot\ for all wind recipes, where the solid red line illustrates the maximum mass for the \citet{SV2020} models. Red triangles correspond to models calculated with SV20 winds, green triangles with H95+, and blue triangles with NL00. Implications of the Alternative Grid on the maximum BH mass across $Z$ with prior evolution taken into consideration. }
    \label{MZ}
\end{figure}

Our findings are summarised in Fig.\,\ref{MZ}, where we depict the maximum final mass formed from our alternative models as a function of the host metallicity, indicating the results using the SV20 recipe with a red line. In all cases, the highest final masses correspond to models with the highest initial He masses.

\section{Discussion on Pair Instability}\label{PPI}
The upper mass limit of stellar mass black holes below the pair instability gap is a key issue for theorists which has multiple uncertainties from massive star evolution, and scarce observational constraints. As the last evolutionary stage before core collapse with a time scale where winds can remove a considerable amount of mass from the star, the mass loss of He stars provides a crucial ingredient that could determine the upper BH mass limit. If indeed, the extent of these WR winds is responsible for setting the upper BH mass limit, then each recipe must be tested for convergence to the point where a maximum final mass is reached irrespective of increasing initial mass. 
In this work, we tested the range of masses produced at core He-exhaustion (see Fig.\,\ref{MfZnew}) finding a critical mass limit for all $Z$ above 2\% \Zdot, i.e. irrespective of initial mass, with the NL00 recipe. Interestingly, this is not the case for the SV20 and H95+ models, which seem to have increased final masses with increased initial masses at all Z. This means that if WR winds were most closely represented by the NL00 prescription, then the upper BH mass limit would indeed be set by WR mass loss. 

\begin{figure}
	\includegraphics[width=\columnwidth]{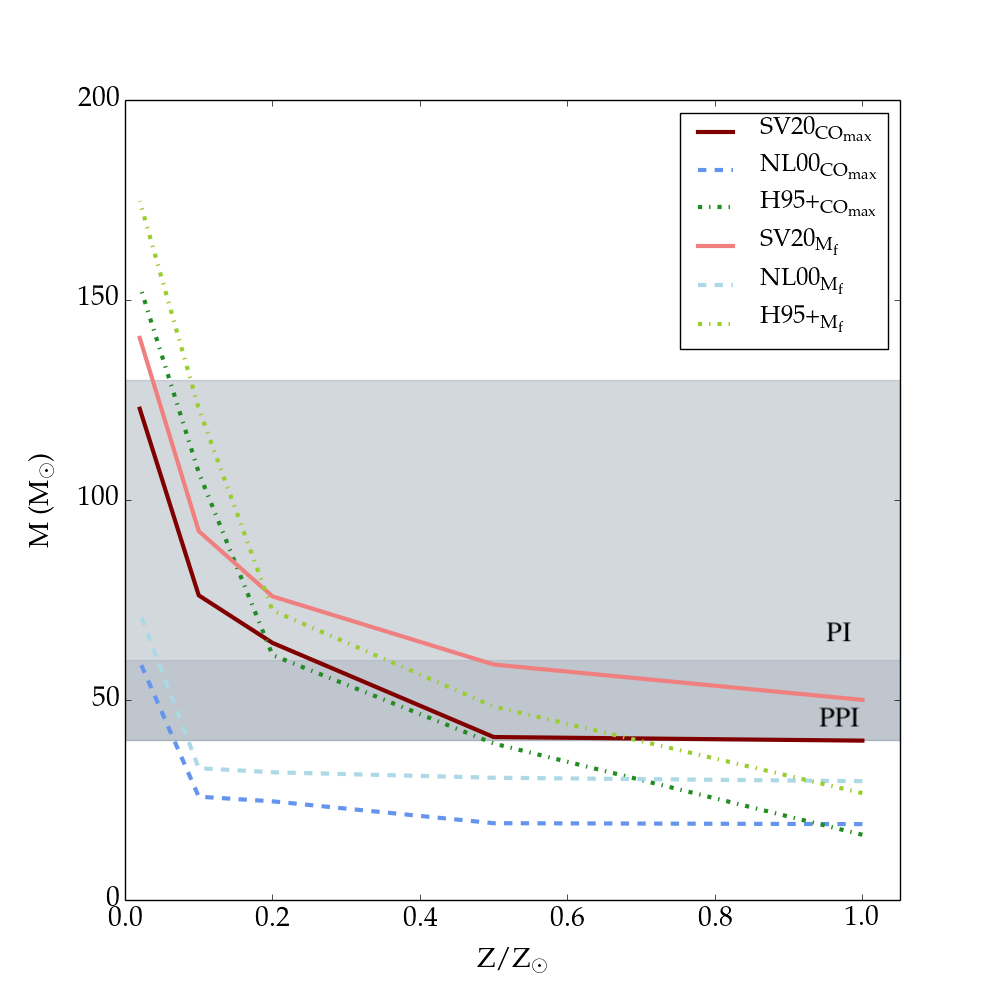}
	\caption{Final CO masses at core He-exhaustion as a function of initial metallicity for each wind recipe. SV20 are represented by solid red lines, H95+ in dashed-dotted green lines and NL00 in dashed blue lines. The maximum final mass (M$_{\rm{f}}$) models, for each metallicity, which include mass-loss rates from SV20 are represented by solid light red lines, H95+ in dashed-dotted light green lines and NL00 in dashed light blue lines. The grey shaded region highlights the region where pulsational pair-instability supernovae occur, with CO core masses above 40\Mdot.}
    \label{CloseupCOcore}
\end{figure}

Since there is no convergence on the upper mass limit by SV20 and H95+ models, such that we may have much higher final masses if extremely high initial masses are invoked, we also must consider other dependencies. Figure \ref{MfMhenew} highlights the correlation of initial He ZAMS mass with the final BH mass, particularly for the SV20 and H95+ models (solid and dashed lines respectively), suggesting that the upper BH mass limit could be set by the maximum initial He ZAMS mass at each Z. When calculating our grid of alternative models, we found that due to strong mass loss on the MS, models with $\sim$ \Zdot\ have reduced He ZAMS masses compared to even \ZSMC\ models. We excluded this effect of the prior evolution on core He-burning in our Standard Grid of models by excluding H-burning mass loss, in order to probe the full WR mass range across all Z. Still, it is important to note that the MS evolution is key in setting the initial mass of He and WR stars - directly impacting the final mass range. If the initial ZAMS mass is not high enough, and mass loss is strong enough on the MS, then our final BH mass limit will be dominated by the prior evolution on the MS. We do not explore the complete picture of pre-WR evolution in this work, as this in itself contains many uncertainties (such as treatment of internal mixing, proximity to the Eddington limit at the highest initial masses, and the extent of MS mass loss). However, we can illustrate the importance of the MS evolution through results from our Alternative Grid (Sect. \ref{QCHEgrid}), where we include the QCHE method as a proxy for evolutionary channels towards WRs. Ultimately a combination of mixing and mass loss will provide various channels towards He stars, and our results show that this can lead to lowered initial He masses, and as a result lower final masses at high $Z$ ($\sim$ 0.2-1\Zdot).

If we consider a wide range of initial He ZAMS masses, but with the realisation that WR winds may not set the upper BH mass limit (as hinted by the recent SV20 results), and the initial mass may drive the maximum final mass up to extremely high final masses, pair instability would introduce an additional cut-off for the resulting BH masses.
Stars above a certain CO core mass do not collapse to form BHs at the end of their lives -- except for a range of very massive stars with M$_{\rm{He}}$ above $\sim$133 \Mdot\ \citep{HW2002} -- but rather rip themselves apart in pair instability supernovae leaving no remnant behind. 

In the lower mass range of this regime pulsation pair instability (PPI) occurs where large eruptions from pulsations lead to the loss of a significant amount of mass. This leads to a so-called PI gap in the resulting BH mass range. This region could be located at varying mass ranges for different Z, directly impacted by the mass loss history of the stars during H and He-burning. Studies by \cite{Woosley2017, Farmer+2019} propose that the mass range above which PPI occurs is driven by the CO core mass. In fact, stars which have a CO core above $\sim$ 40\Mdot\ are predicted to enter the PPI gap, \citep{Woosley19, Woosley+2020}. In order to have increased final masses which do not enter the PPI regime, stars should have small convective cores. \cite{vink2020} shows that by maintaining a small core with low convective core overshooting (\fov $=$ 0.01), and retaining a large H envelope, the effective PI gap can be much smaller than commonly anticipated as even massive BHs up to about 90 \Mdot, such as the one seen in the GW event GW190521, may be formed.

In this study, we endeavour to calculate grids of WR models through various methods, mainly by increased internal mixing through convection (Standard Grid) and rotation (Alternative Grid). Higher internal mixing leads to increased convective core masses, and as a result will push our WR models towards PPI at lower masses, since the core masses in our models are relatively high, compared to models with standard/lowered mixing (e.g. \aov $\sim$ 0.1-0.5, or $<$40\% critical rotation). Figure \ref{CloseupCOcore} highlights the small difference between the CO core mass and total final mass (e.g. see change from red line to pink line, or light blue line to blue line) for all $Z$ and WR mass-loss prescriptions. 
At \Zdot\ and \ZLMC, most of our calculations predict BHs of varying mass depending on the wind prescription. This situation alters at lower Z, where even at SMC, more PPISN are predicted, and for lower initial masses $-$ from 200\Mdot\ down to 100\Mdot\ for SV20 and H95+ models. At 2-10\% \Zdot, PI already occurs for 60\Mdot\ He-star models and higher, again with rates from SV20 and H95+. NL00 models only enter PI at 2\% \Zdot\ with initial He masses $\geq$60\Mdot. 

The critical upper $Z$ limit for the occurrence of pair instability, $Z_{\rm{PI}}$, also sets the limit of maximum BH masses and seems to be in the 10-50\% \Zdot\ region. The higher $Z$ models are set by the initial He mass, dominated by prior evolution on the MS. At the lowest $Z$ (2-10\% \Zdot) all three recipes predict PPISN, but only SV20 and H95+ models predict PPISN at 0.02 $<$  $Z$/\Zdot $<$0.5, giving a much wider $Z$ range. Interestingly, the observed $Z_{\rm{PI}}$ could be lower than the derived $Z_{\rm{PI}}$ due to prior evolution limiting the range of He star masses. Based on our calculations approximating prior evolution, we would expect to have the observed $Z_{\rm{PI}}$ at $Z$ $\leq$ 0.1 \Zdot.
These results will  have consequences for observations of PPISN/PISN and the spectrum of BH masses at varied Z.

\section{Conclusions}\label{conclusions}
In this study we have presented two grids of stellar evolution models which incorporate the mass-loss description from \cite{SV2020} based on dynamically-consistent atmosphere calculations of massive helium star winds. Our evolution models span a wide range of initial masses and metallicities, focusing on the core He-burning phase. As a comparison we have calculated models with standard hot, helium star winds as implemented in many stellar evolution codes from \cite{NugisLamers}, as well as models which include \cite{Yoon06} rates commonly implemented in population synthesis codes \citep[e.g.][]{Belczynski+2010}.
The change in final mass as a function of wind recipe highlights the importance of implementing proper descriptions of wind mass loss in stellar evolution codes. 
By applying the first theoretically derived mass-loss description for classical Wolf-Rayet stars (SV20), we included a qualitative and quantitative improvement to stellar evolution models of He stars, which also provides an insight into the range of black hole masses at various metallicities. This qualitative difference in the treatment of WR winds can be observed in the behaviour of the upper BH mass limit at each $Z$. Our results uncover the convergence of a maximum BH mass with the NL00 prescription, whereas there is no convergence on the final mass for models implementing the SV20 description. As a result, our grids illustrate that if there is no convergence based on wind recipes then the prior evolution or pair instability is responsible for limiting the upper BH mass. 

At solar metallicity, the change in the final mass due to implementing the SV20 description is generally limited. Nonetheless, the SV20 description allows for higher final masses at \Zdot\ than other descriptions do. Due to the more complex shape of the description by SV20, this is not true at the lowest $Z$-range, where the H95+ description would allow slightly higher final masses than SV20. 
At all Z, models which implement the NL00 recipe have final masses below $\sim$ 30\Mdot, except for the highest masses at 0.02\Zdot\ shown in Sect. \ref{newgridSect}.
This means that a 40\Mdot\ BH could not be formed from a WR star using the standard 'Dutch' wind recipe as included in MESA, unless an extremely low $Z$ is enforced. 
In contrast, we find that with the SV20 mass loss a 40\Mdot\ BH may be formed in the range 0.1-0.2 \Zdot, with final masses up to 56\Mdot\ at 0.1\Zdot. While we did not continue our calculations after encountering PI, we do find interesting limitations. Our results highlight that the NL00 wind prescription leads to no stellar BHs formed above the second BH mass gap, unlike with SV20. Extremely low metallicity environments such as I Zw 18 ($\sim 0.02\,$\Zdot) provide an insight into the early Universe and likely host some of the heaviest stellar mass black holes which may be detected as gravitational wave sources \citep{vink2020}. Calculations with the SV20 description show final masses of up to $140$\Mdot\ at $0.02$\Zdot, hinting that there might be a channel for very massive He stars to form black holes above the second BH mass gap, though the most massive BHs may form from H-rich stars. 

We have analysed the dependencies on the critical mass limit of BHs below the pair instability gap for a range of $Z$. Our calculations show that only the standard, self-enriched NL00 mass-loss rates really yield a limit due to WR mass loss, in this case at 20-30\Mdot, while the new physically-motivated SV20 rates, and H95+ models do not converge towards such a critical mass limit. Thus, the initial mass of the He star is the determining factor. This value is both influenced by the prior evolution on the MS (including the H ZAMS mass) and the effects of mass loss before reaching the He ZAMS.
Eventually, PPI effectively avoids the formation of BHs above a certain CO core mass limit. Interestingly, the upper BH mass for each $Z$ can also be used as a proxy for the Z$_{\rm{PI}}$, where pair instability supernovae may be observed. With the NL00 mass loss prescription, only models at (and below) 2\% \Zdot\ eventually enter the PI regime. This Z$_{\rm{PI}}$ limit is increased by models with the SV20 and H95+ descriptions, reaching this PI regime already at 10-50\% \Zdot, depending on what we consider as a reasonable upper limit for the He star mass. Our results have consequences for the observations of (P)PISNe, which could already occur at considerably higher metallicities than previously assumed, and the mass spectrum of observed BHs as a function of $Z$. 

\section*{Acknowledgements}
This article is based upon work from the “ChETEC” COST Action (CA16117), supported
by COST (European Cooperation in Science and Technology)”. RH acknowledges support from the IReNA AccelNet Network of Networks, supported by the National Science Foundation under Grant No. OISE-1927130 and from the World Premier International Research Centre Initiative (WPI Initiative), MEXT, Japan. The authors would like to thank Sarah Lightfoot for preliminary test models during her Ogden Trust Summer internship at Keele in 2019. JSV and ERH are supported by STFC funding under grant number ST/R000565/1 in the context of the BRIDGCE UK Network. AACS is \"{O}pik Research Fellow at Armagh Observatory \& Planetarium.
\section*{Data Availability}

The data underlying this article will be shared on reasonable request
to the corresponding author.
\typeout{}
\bibliographystyle{mnras}
\bibliography{newdiff} 

\appendix
 \section{Tables of model grids and additional figures}

\begin{table}
    \centering
    \caption{Standard Grid of Helium models for Sect. \ref{newgridSect} with zero mass loss during core H-burning, for \Zdot, 0.5\Zdot\ and 0.2\Zdot.  Final masses are provided at core He-exhaustion, as well as the final CO core masses.}
    \begin{tabular}{c|c|c|c|c}
    \hline
       $\dot{M}_\mathrm{recipe}$ & Z/\Zdot & $M_{\rm{ini}}$ & $M_{\rm{f}}$ & $M_{\rm{CO}}$ \\
       \hline\hline
        SV20 & 1 & 20 & 12.51 & 9.65\\
        SV20 & 1 & 40 & 17.81 & 14.47\\
        SV20 & 1 & 60 & 22.46 & 18.80\\
        SV20 & 1 & 80 & 26.46 & 22.90\\
        SV20 & 1 & 100 & 30.99 & 26.83\\
        SV20 & 1 & 120 & 34.97 & 30.63\\
        SV20 & 1 & 200 & 50.03 & 39.84\\
        \hline
        NL00 & 1 & 20 & 11.04 & 8.25\\
        NL00 & 1 & 40 & 15.04 & 11.87\\
        NL00 & 1 & 60 & 18.04 & 14.62\\
        NL00 & 1 & 80 & 20.49 & 16.89\\
        NL00 & 1 & 100 & 22.56 & 18.80\\
        NL00 & 1 & 120 & 24.47 & 20.51\\
        NL00 & 1 & 200 & 29.71 & 25.32\\
        \hline
        H95+ & 1 & 20 & 11.60 & 8.79\\
        H95+ & 1 & 40 & 16.33 & 13.12\\
        H95+ & 1 & 60 & 19.23 & 15.81\\
        H95+ & 1 & 80 & 21.29 & 17.80\\
        H95+ & 1 & 100 & 22.80 & 19.13\\
        H95+ & 1 & 120 & 23.99 & 20.32\\
        H95+ & 1 & 200 & 26.87 & 23.07\\
        \hline\hline
        SV20 & 0.5 & 20 & 15.45 & 12.33\\
        SV20 & 0.5 & 40 & 22.04 & 18.31\\
        SV20 & 0.5 & 60 & 27.57 & 23.55\\
        SV20 & 0.5 & 80 & 32.80 & 28.41\\
        SV20 & 0.5 & 100 & 37.56 & 32.84\\
        SV20 & 0.5 & 120 & 41.85 & 36.90\\
        SV20 & 0.5 & 200 & 58.87 & 40.74\\
        \hline
        NL00 & 0.5 & 20 & 12.45 & 9.50\\
        NL00 & 0.5 & 40 & 17.25 & 13.85\\
        NL00 & 0.5 & 60 & 19.90 & 16.34\\
        NL00 & 0.5 & 80 & 22.08 & 18.34\\
        NL00 & 0.5 & 100 & 24.40 & 20.50\\
        NL00 & 0.5 & 120 & 23.54 & 19.66\\
        NL00 & 0.5 & 200 & 30.57 & 26.14\\
        \hline
        H95+ & 0.5 & 20 & 14.25 & 11.19\\
        H95+ & 0.5 & 40 & 22.41 & 18.65\\
        H95+ & 0.5 & 60 & 28.29 & 24.09\\
        H95+ & 0.5 & 80 & 32.98 & 28.48\\
        H95+ & 0.5 & 100 & 36.67 & 32.00\\
        H95+ & 0.5 & 120 & 39.54 & 34.57\\
        H95+ & 0.5 & 200 & 48.37 & 39.12\\
        \hline
        \hline
        SV20 & 0.2 & 20 & 19.25 & 16.01\\
        SV20 & 0.2 & 40 & 28.31 & 24.07\\
        SV20 & 0.2 & 60 & 35.60 & 30.80\\
        SV20 & 0.2 & 80 & 42.22 & 37.08\\
        SV20 & 0.2 & 100 & 48.45 & 42.86\\
        SV20 & 0.2 & 120 & 54.34 & 44.52\\
        SV20 & 0.2 & 200 & 75.91 & 64.28\\
        \hline
        NL00 & 0.2 & 20 & 12.63 & 9.66\\
        NL00 & 0.2 & 40 & 19.91 & 16.31\\
        NL00 & 0.2 & 60 & 22.06 & 18.29\\
        NL00 & 0.2 & 80 & 24.35 & 20.38\\
        NL00 & 0.2 & 100 & 24.88 & 20.92\\
        NL00 & 0.2 & 120 & 27.88 & 23.69\\
        NL00 & 0.2 & 200 & 31.93 & 27.33\\
        \hline
    \end{tabular}
    \label{newzeroMdot}
\end{table}

\begin{table}
    \centering
    \caption{Continued. Standard Grid of models, for 0.2\Zdot, 0.1\Zdot\ and 0.02\Zdot.}
    \begin{tabular}{c|c|c|c|c}
    \hline
      $\dot{M}_\mathrm{recipe}$ & Z/\Zdot & $M_{\rm{ini}}$ & $M_{\rm{f}}$ & $M_{\rm{CO}}$ \\
      \hline\hline
        H95+ & 0.2 & 20 & 16.02 & 12.78\\
        H95+ & 0.2 & 40 & 27.09 & 23.00\\
        H95+ & 0.2 & 60 & 35.92 & 31.14\\
        H95+ & 0.2 & 80 & 43.28 & 37.99\\
        H95+ & 0.2 & 100 & 49.70 & 44.04\\
        H95+ & 0.2 & 120 & 55.34 & 45.57\\
        H95+ & 0.2 & 200 & 72.38 & 61.23\\
        \hline\hline
        SV20 & 0.1 & 20 & $-$ & $-$\\
        SV20 & 0.1 & 40 & 32.96 & 28.33\\
        SV20 & 0.1 & 60 & 41.75 & 36.51\\
        SV20 & 0.1 & 80 & 49.84 & 44.01\\
        SV20 & 0.1 & 100 & 57.48 & 47.18\\
        SV20 & 0.1 & 120 & 65.10 & 53.88\\
        SV20 & 0.1 & 200 & 92.22 & 76.16\\
        \hline
        NL00 & 0.1 & 20 & 16.71 & 13.45\\
        NL00 & 0.1 & 40 & 26.52 & 22.41\\
        NL00 & 0.1 & 60 & 30.22 & 25.82\\
        NL00 & 0.1 & 80 & 32.18 & 27.61\\
        NL00 & 0.1 & 100 & 32.50 & 27.93\\
        NL00 & 0.1 & 120 & 27.76 & 23.58\\
        NL00 & 0.1 & 200 & 32.95 & 28.31\\
        \hline
        H95+ & 0.1 & 20 & 18.16 & 15.41\\
        H95+ & 0.1 & 40 & 33.60 & 28.95\\
        H95+ & 0.1 & 60 & 47.54 & 41.61\\
        H95+ & 0.1 & 80 & 60.37 & 43.87\\
        H95+ & 0.1 & 100 & 72.32 & 60.02\\
        H95+ & 0.1 & 120 & 83.77 & 68.80\\
        H95+ & 0.1 & 200 & 122.87 & 106.88\\
        \hline\hline
        SV20 & 0.02 & 20 & $-$  & $-$\\
        SV20 & 0.02 & 40 & $-$  & $-$\\
        SV20 & 0.02 & 60 & $-$  & $-$\\
        SV20 & 0.02 & 80 & 74.09 & 63.27\\
        SV20 & 0.02 & 100 & 85.52 & 71.00\\
        SV20 & 0.02 & 120 & 96.86 & 82.96\\
        SV20 & 0.02 & 200 & 140.53 & 122.81\\
        \hline
        NL00 & 0.02 & 20 & 13.46 & 10.47\\
        NL00 & 0.02 & 40 & 35.40 & 30.79\\
        NL00 & 0.02 & 60 & 47.82 & 41.83\\
        NL00 & 0.02 & 80 & 56.44  & 46.35\\
        NL00 & 0.02 & 100 & 62.13 & 50.28\\
        NL00 & 0.02 & 120 & 66.56 & 46.86\\
        NL00 & 0.02 & 200 & 71.75 & 59.62\\
        \hline
        H95+ & 0.02 & 20 & 19.51 & 16.03\\
        H95+ & 0.02 & 40 & 38.20 & 34.28\\
        H95+ & 0.02 & 60 & 56.35  & 44.43\\
        H95+ & 0.02 & 80 & 74.11 & 63.41\\
        H95+ & 0.02 & 100 & 91.54 & 76.72\\
        H95+ & 0.02 & 120 & 108.72 & 89.51\\
        H95+ & 0.02 & 200 & 174.75 & 153.65\\
        \hline
        \hline
    \end{tabular}
    \label{newzeroMdot2}
\end{table}

\begin{table}\caption{Alternative Grid of models for a range of initial H and He masses (calculated where X$_{c}$ $<$ 0.00001) as a function of initial $Z$ and wind recipe. Final masses are provided at core He-exhaustion, as well as the final CO core masses.}
    \centering
    \begin{tabular}{|c|c|c|c|c|c|}
        \hline
         $\dot{M}_\mathrm{recipe}$ & $Z/$\Zdot & $M_{\rm{H}}$ & $M_{\rm{He}}$ & $M_{\rm{f}}$ & $M_{\rm{CO}}$ \\
         \hline
            SV20 & 1 & 30 & 10.784 & 9.301 & 6.821 \\
            SV20 & 1 & 50 & 12.907 & 10.153 & 7.495 \\
            SV20 & 1 & 70 & 15.040 & 10.887 & 8.165 \\
            SV20 & 1 & 100 & 17.620 & 11.742 & 8.962 \\
            SV20 & 1 & 120 & 18.965 & 12.162 & 9.312 \\
            SV20 & 1 & 150 & 21.099 & 12.806 & 9.893 \\
            SV20 & 1 & 170	& 22.101 & 13.111 & 10.160 \\
            \hline
            NL00 & 1 & 30 & 10.771 & 7.669 & 5.296 \\
            NL00 & 1 & 50 & 12.889 & 8.768 & 6.331 \\
            NL00 & 1 & 70 & 15.038 & 9.072 & 6.510 \\
            NL00 & 1 & 100 & 17.619 & 10.066 & 7.399 \\
            NL00 & 1 & 120	& 18.965 & 10.627 & 7.912 \\
            NL00 & 1 & 150	& 21.101 & 11.256 & 8.508 \\
            NL00 & 1 & 170 & 22.103 & 11.456 & 8.638 \\
            \hline
            H95+ & 1 & 30	& 10.772 & 7.837 & 5.445\\
            H95+ & 1 & 50	& 12.888 & 8.912 & 6.387\\
            H95+ & 1 & 70	& 15.038 & 9.780 & 7.160 \\
            H95+ & 1 & 100	& 17.616 & 10.764 & 8.046 \\
            H95+ & 1 & 120	& 18.961 & 11.231 & 8.470 \\
            H95+ & 1 & 150	& 21.095 & 11.918 & 9.114 \\
            H95+ & 1 & 170	& 22.097 & 12.233 & 9.372\\
            \hline
            \hline
            SV20 & 0.5 & 30 & 13.572 & 11.456 & 10.044 \\
            SV20 & 0.5 & 50 & 18.797 & 14.773 & 11.698 \\
            SV20 & 0.5 & 70 & 21.316 & 15.796 & 12.655 \\
            SV20 & 0.5 & 100 & 25.699 & 17.397 & 14.058 \\
            SV20 & 0.5 & 120 & 29.575 & 18.724 & 15.270\\
            SV20 & 0.5 & 150 & 32.614 & 19.763 & 16.225 \\
            SV20 & 0.5 & 170 & 34.610 & 20.382 & 16.824 \\
            \hline
            NL00 & 0.5 & 30 & 13.572 & 9.914 & 7.296 \\
            NL00 & 0.5 & 50 & 18.788 & 11.575 & 8.724 \\
            NL00 & 0.5 & 70 & 21.315 & 12.871 & 9.896 \\
            NL00 & 0.5 & 100 & 25.697 & 14.177 & 11.123 \\
            NL00 & 0.5 & 120 & 29.577 & 15.104 & 11.929 \\
            NL00 & 0.5 & 150 & 32.620 & 15.848 & 12.592 \\
            NL00 & 0.5 & 170 & 34.610 & 16.097 & 12.873 \\
            \hline
            H95+ & 0.5 & 30 & 13.572 & 10.484 & 7.858 \\
            H95+ & 0.5 & 50 & 18.790 & 13.407 & 10.473 \\
            H95+ & 0.5 & 70 & 21.315 & 14.752 & 11.681 \\
            H95+ & 0.5 & 100 & 25.697 & 16.821 & 13.513 \\
            H95+ & 0.5 & 120 & 29.576 & 18.484 & 15.059 \\
            H95+ & 0.5 & 150 & 32.616 & 19.756 & 16.254 \\
            H95+ & 0.5 & 170 & 34.610 & 20.495 & 16.904 \\
            \hline
            \hline 
            SV20 & 0.2 & 30 & 16.074 & 16.070 & 12.857 \\
            SV20 & 0.2 & 50 & 23.581 & 20.824 & 17.317 \\
            SV20 & 0.2 & 70 & 30.854 & 24.145 & 20.289 \\
            SV20 & 0.2 & 100 & 37.014 & 26.804 & 22.695 \\
            SV20 & 0.2 & 120 &  40.170 & 28.124 & 23.983\\
            SV20 & 0.2 & 150 & 51.356 & 32.514 & 27.999 \\
            SV20 & 0.2 & 170 & 51.551 & 32.613 & 28.121 \\
            \hline
        
    \end{tabular}
    \label{masstable}
\end{table}

\begin{table}
    \centering
    \caption{Continued. Alternative Grid of models, for 0.2\Zdot, 0.1\Zdot\ and 0.02\Zdot.}
    \begin{tabular}{|c|c|c|c|c|c|}
        \hline
         $\dot{M}_\mathrm{recipe}$ & $Z/$\Zdot & $M_{\rm{H}}$ & $M_{\rm{He}}$ & $M_{\rm{f}}$ & $M_{\rm{CO}}$ \\
         \hline
            NL00 & 0.2 & 30 & 16.066 & 11.584 & 9.048\\
            NL00 & 0.2 & 50 & 23.554 & 14.611 & 11.548 \\
            NL00 & 0.2 & 70 & 30.854 & 16.637 & 13.352 \\
            NL00 & 0.2 & 100 & 37.007 & 18.328 & 14.939 \\
            NL00 & 0.2 & 120 & 40.164 & 19.071 & 15.547 \\
            NL00 & 0.2 & 150 & 51.357 & 20.848 & 17.219 \\
            NL00 & 0.2 & 170 & 51.551 & 21.493 & 17.812\\
            \hline
            H95+ & 0.2 & 30 & 16.068 & 12.995 & 10.301\\
            H95+ & 0.2 & 50 & 23.562 & 17.770 & 14.421 \\
            H95+ & 0.2 & 70 & 30.854 & 21.874 & 18.205 \\
            H95+ & 0.2 & 100 & 37.010 & 25.268 & 21.288 \\
            H95+ & 0.2 & 120 & 40.167 & 26.911 & 22.841 \\
            H95+ & 0.2 & 150 & 51.357 & 32.243 & 27.724 \\
            H95+ & 0.2 & 170 & 51.551 & 32.407 & 28.001 \\
            \hline
            \hline
            SV20 & 0.1 & 30 & 20.886 & 20.886 & 16.919 \\
            SV20 & 0.1 & 50 & 34.844 & 28.807 & 24.855 \\
            SV20 & 0.1 & 70 & 8.436 & 33.884 & 29.563 \\
            SV20 & 0.1 & 100 & 82.126 & 42.391 & 37.410 \\
            SV20 & 0.1 & 120 & 94.747 & 46.770 & 41.181\\
            SV20 & 0.1 & 150 & 101.778 & 54.201 & $-$ \\
            SV20 & 0.1 & 170 & 100.529 & 55.786 & $-$ \\
            \hline
            NL00 & 0.1 & 30 & 20.864 & 13.702 & 10.837 \\
            NL00 & 0.1 & 50 & 34.843 & 13.017 & 10.085 \\
            NL00 & 0.1 & 70 & 48.436 & 16.431 & 13.224 \\
            NL00 & 0.1 & 100 & 82.055 & 17.219 & 13.916 \\
            NL00 & 0.1 & 120 & 94.843 & 19.228 & 15.738 \\
            NL00 & 0.1 & 150 & 101.708 & 25.936 & 21.926 \\
            NL00 & 0.1 & 170 & 100.536 & 31.827 & 27.307\\
            \hline
            H95+ & 0.1 & 30 & 20.875 & 17.916 & 14.409 \\
            H95+ & 0.1 & 50 & 34.844 & 27.269 & 23.006 \\
            H95+ & 0.1 & 70 & 48.439 & 35.903 & 31.083 \\
            H95+ & 0.1 & 100 & 82.213 & 49.726 & 43.747 \\
            H95+ & 0.1 & 120 & 94.725 & 57.530 & $-$ \\
            H95+ & 0.1 & 150 & 101.882 & 68.791 & $-$ \\
            H95+ & 0.1 & 170 & 100.539 & 70.527 & $-$ \\	
            \hline
            \hline
            SV201 & 0.02 & 50 & 44.000 & 44.000 & 37.443 \\		
            SV20 & 0.02 & 100 & 79.529 & 65.719 & $-$ \\
            SV20 & 0.02 & 120 & 89.442 & 71.796 & $-$ \\
            SV20 & 0.02 & 150 & 140.635 & 91.535 & $-$ \\
            SV20 & 0.02 & 170 & 155.078 & 99.927 & $-$ \\
            \hline
            NL00 & 0.02 & 30 & 25.658 & 10.994 & 8.299 \\
            NL00 & 0.02 & 50 & 44.272 & 11.415 & 8.669 \\
            NL00 & 0.02 & 70 & 62.588 & 14.701 & 11.661 \\
            NL00 & 0.02 & 100 & 79.491 & 18.825 & 15.358 \\
            NL00 & 0.02 & 120 & 89.442 & 23.006 & 19.262 \\
            NL00 & 0.02 & 150 & 140.228 & 22.461 & 18.694 \\
            NL00 & 0.02 & 170 & 155.138 & 27.149 & 23.083 \\
            \hline
            H95+ & 0.02 & 30 & 25.661 & 21.938 & 18.198 \\
            H95+ & 0.02 & 50 & 44.418 & 32.343 & 28.744 \\
            H95+ & 0.02 & 70 & 62.588 & 45.752 & 40.418 \\
            H95+ & 0.02 & 100 & 79.520 & 62.506 & $-$ \\
            H95+ & 0.02 & 120 & 89.442 & 73.388 & $-$ \\
            H95+ & 0.02 & 150 & 140.426 & 104.398 & $-$ \\
            H95+ & 0.02 & 170 & 155.209 & 117.140 & $-$ \\         
          \hline
    \end{tabular}
    \label{tab1}
\end{table}

\begin{figure}
	\includegraphics[width=\columnwidth]{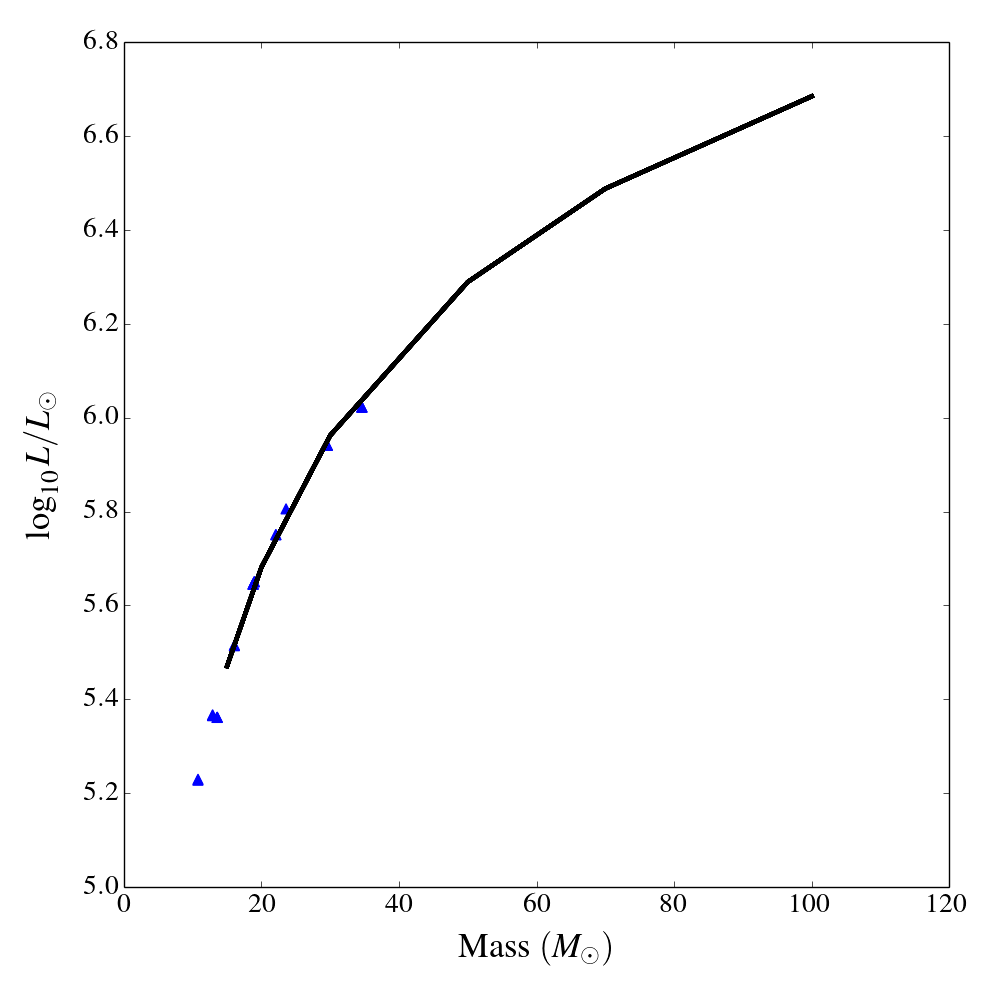}
    \caption{Mass-Luminosity relation for He ZAMS models with SV20 mass-loss rates (blue triangles), with a comparison of the M-L relation derived from G+11 for pure helium stars (black solid line).}
    \label{GotzML}
\end{figure}

\begin{figure}
    \centering
    \includegraphics[width=\columnwidth]{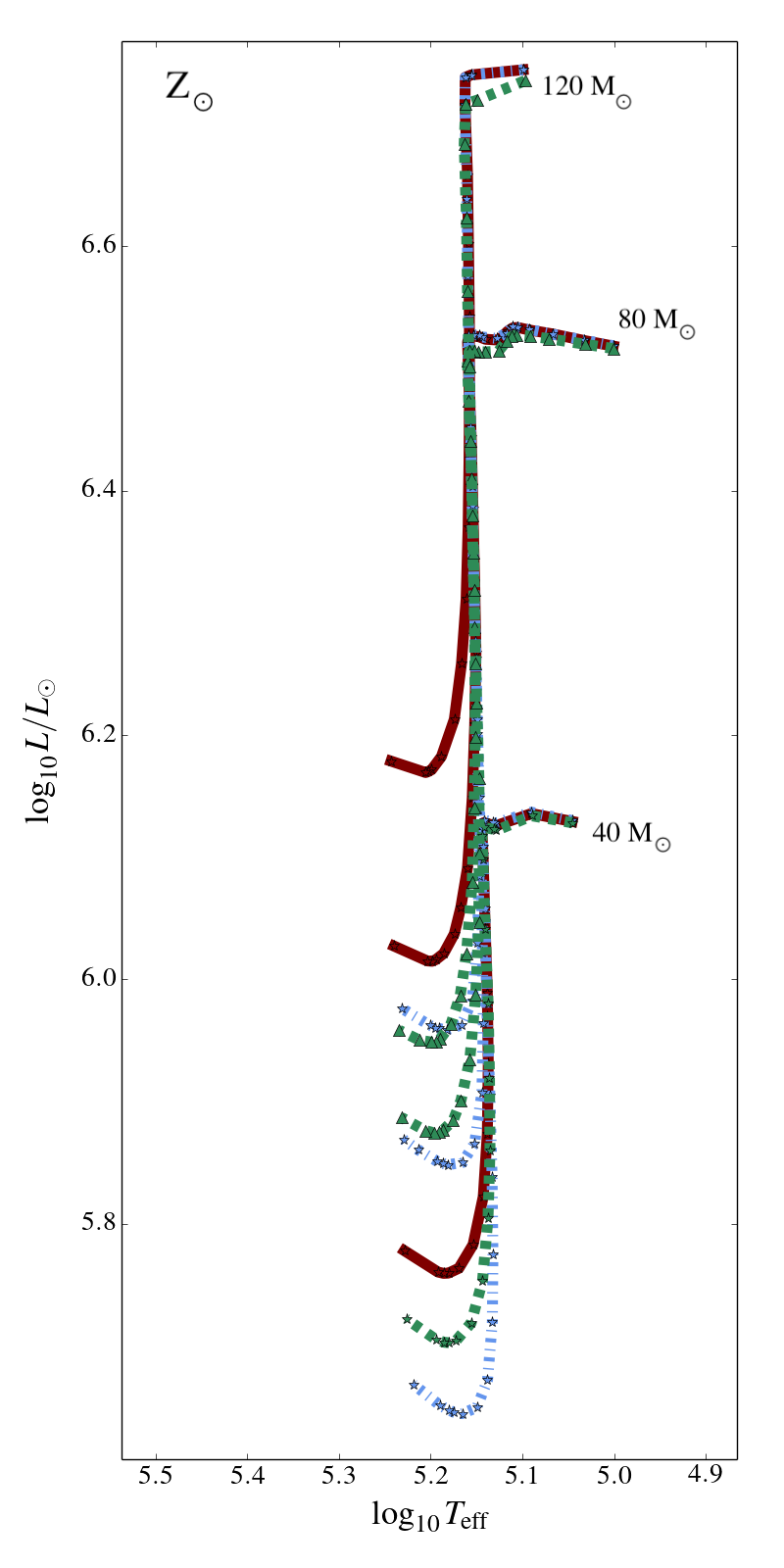}
    \caption{Hertzsprung-Russell diagram of 40-120\Mdot\ He-burning stars evolving from the He-ZAMS as Wolf-Rayets, calculated at \Zdot. Red solid tracks represent models which employ SV20 mass loss, those with NL00 rates are shown in blue dash-dotted lines, and green dashed lines illustrate models which include H95+ mass loss. }
    \label{HRD2solar}
\end{figure}

\begin{figure}
    \centering
    \includegraphics[width=\columnwidth]{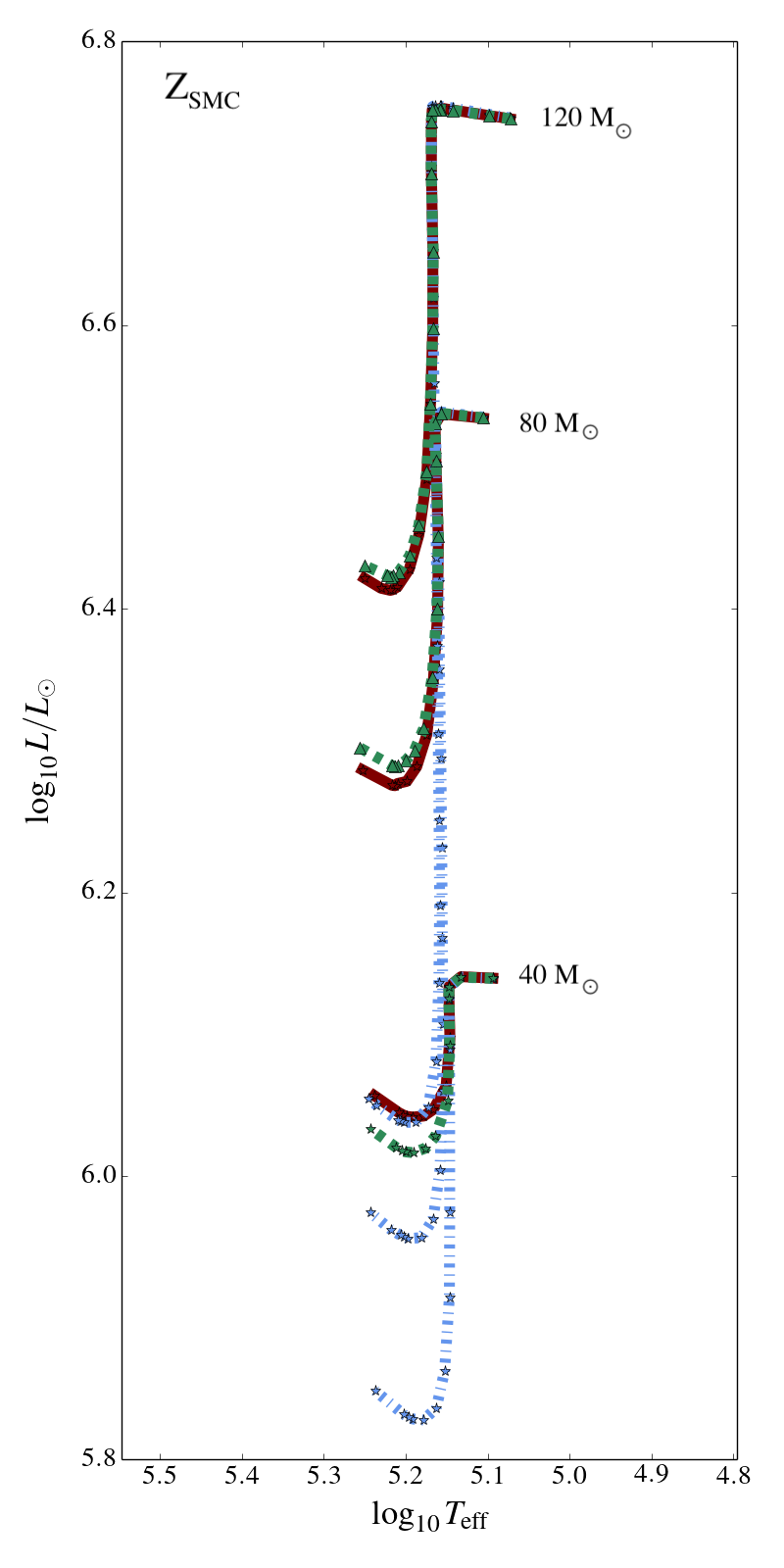}
    \caption{Hertzsprung-Russell diagram of 40-120\Mdot\ He-burning stars evolving from the He-ZAMS as Wolf-Rayets, calculated at \ZSMC. Red solid tracks represent models which employ SV20 mass loss, those with NL00 rates are shown in blue dash-dotted lines, and green dashed lines illustrate models which include H95+ mass loss. }
    \label{HRD2smc}
\end{figure}

\bsp	
\label{lastpage}
\end{document}